\documentclass[useAMS,usenatbib]{mn2e}
\usepackage{graphicx}
\usepackage{epsfig}
\usepackage{amssymb}
\usepackage{lscape}
\usepackage{ulem}
\usepackage{txfonts}
\usepackage{lipsum}
\usepackage{float}
\usepackage{hyperref}

\urldef\myurl\url{https://wwwmpa.mpa-garching.mpg.de/SDSS/DR7/}

\def\s21{S$_{21}$}
\def\ha{H$\alpha$}
\def\hb{H$\beta$}
\def\nha{[NII]/H$\alpha$}
\def\ohb{[OIII]/H$\beta$}

\title[SF versus AGN classifications: A machine learning approach ] {The discrimination  between star-forming  and AGN galaxies in the absence of \ha\ and [NII]: A  machine learning approach}

\author[H. Teimoorinia and J. Keown] {Teimoorinia, H.$^{1}$ and Keown, J. $^2$ \\ 
$^1$ NRC Herzberg Astronomy and Astrophysics, 5071 West Saanich Road, Victoria, BC, V9E 2E7, Canada \\
$^2$ Department of Physics and Astronomy, University of Victoria, Victoria, BC, V8P 5C2, Canada\\}

\def\LaTeX{L\kern-.36em\raise.3ex\hbox{a}\kern-.15em
    T\kern-.1667em\lower.7ex\hbox{E}\kern-.125emX}

\begin{document}

\label{firstpage}

\maketitle

\begin{abstract}

In the absence of the two emission lines \ha\ and [NII] (6584\AA) in a BPT diagram,  we show that other spectral information is sufficiently informative to distinguish AGN galaxies from star-forming galaxies. We use pattern recognition methods and  a sample of galaxy spectra from the Sloan Digital Sky Survey (SDSS)  to  show that, in this survey,  the flux and equivalent width of [OIII] (5007\AA) and H$\beta$, along with the 4000\AA ~break, can be used to classify galaxies in a BPT diagram. This method provides a higher accuracy of predictions than those which use stellar mass and \ohb. First, we use BPT diagrams and various physical parameters to re-classify the galaxies.  Next, using confusion matrices, we determine  the `correctly' predicted classes as well as confused cases.  In this way, we investigate the effect of each parameter in the confusion matrices and rank the physical parameters used in the discrimination of the different classes.  We show that in this survey, for example,  $\rm{g - r}$  colour can provide the same  accuracy as galaxy stellar mass to predict whether or not a galaxy hosts an AGN.  Finally, with the same information, we also rank the parameters involved in the discrimination of Seyfert and LINER galaxies.

\end{abstract}


\section{Introduction}
\label{introduction}
The information coming from galaxies can be captured in a combination of photometric and spectroscopic data that occur in different intervals along the electromagnetic spectrum, from X-ray to radio wavelengths. Generally, not all the desired spectroscopic and photometric information are available for an astronomical problem and, in most cases,  we have access to only parts of the information. Each part can have useful clues, so with limited information, it is important to use powerful methods to find informative parameters  and  obtain conclusive results from the available data.  In this respect and as an example, when the emission lines [NII] (6584\AA) and \ha\ are not available in galaxy spectra, discrimination  between star-forming (SF) and AGN galaxies can be a challenging problem. 

Although stars are an important source of energy production in galaxies, non-stellar energy sources can also exist in galaxies.  One such important non-stellar source is related to active galactic nuclei (AGN), which can be found in the centers of most galaxies \citep{Richstone_1998}.  These galaxies  are the most luminous objects in the universe and contain  an accretion disk  surrounding a black hole in their centers. AGN can also be categorized into several sub-classes, such as Seyferts and LINERs, for which the latter group is dominated by older stellar populations that have higher velocity dispersions with respect to the former (\citealt[][hereafter K06]{Kewley_2006}; \citealt{Singh_2013}).  The impact an AGN has upon the evolution of a galaxy, as well as the different mechanisms that create Seyferts versus LINERs, are still heavily debated topics.  Accurately separating star-forming and AGN galaxies based on their different physical and spectral characteristics has been a critical step towards better understanding AGN activity since it has provided the large samples necessary to compare these distinct galaxy classes.  Several methods have been developed to classify galaxies as either AGN or star-forming based on their spectral differences.  For instance, AGN are generally the source of `harder' ionization than the hot stars of star-forming galaxies can provide \citep[][hereafter K01]{Kewley_2001}.  In this context, one of the most popular galaxy classification methods is presented by \cite{Baldwin_1981}; hereafter BPT. This work has a theoretical base in which a set of nebular emission lines, (e.g., [NII], \ha\ , [OIII], and \hb) are used to distinguish the ionization mechanism of nebular gas.  In this way,  based on different classifications,  one can distinguish  a  SF galaxy from an AGN galaxy (e.g., K01 and \citealt{Stasinska_2006}; hereafter S06).  An intermediate class between the two aforementioned classes called `composite' galaxies also exists. There are certain sub-classes of AGN and SF galaxies that do not appear in BPT diagrams, however, including X-ray bright AGN \citep{Winter_2009} and dusty SF galaxies \citep{Rosario_2016}.  In this paper, AGN and SF galaxies  are defined to be above the cutoff line designated by K01 and under the cutoff line from S06, respectively.  Any galaxy falling in-between those lines is defined as a composite galaxy.

To construct a BPT diagram, it is necessary to obtain  `relatively'  high sensitivity measurements (e.g., SNR $>3$) for the [NII], H$\alpha$, [OIII], and H$\beta$ emission lines.   These limitations make it difficult to distinguish  a SF galaxy from an AGN  galaxy in  noisy spectra or  for higher redshift galaxies (i.e., $0.4<z<1$) where the [NII] and \ha\ emission lines can be  absent in optical spectra.  In the absence of near-infrared spectroscopy at intermediate redshifts, \cite{Juneau_2011} (hereafter Ju11) use  \ohb\ and stellar mass for a sample of galaxies and  predict the class of a  galaxy on a BPT diagram. With an alternative approach using Artificial Neural Networks (ANNs) and different non-linear models, \cite{Tem_2014} (henceforth TE14) also show that the [NII] and \ha\ emission lines can be well-predicted using different parameters, such as stellar mass, H$\beta$, and [OIII]. Additionally, TE14 use the predicted emission lines to construct a BPT diagram and show that, on average, more than 86 \% of AGN galaxies can  be correctly classified. TE14, however, do not directly use BPT diagrams to classify galaxies.  Such diagrams can provide more detailed and valuable information such as the probability for a given galaxy to be AGN versus star-forming after classification is completed.

In addition to stellar mass and \ohb , other combinations of physical or spectral parameters have been proposed as effective galaxy diagnostic tools.  For instance, rest-frame colors \citep[the CEx method;][]{Yan_2011}, HeII(4686\AA)/\hb\ versus [N II]/\ha\ \citep{Shirazi_2012}, and the 4000\AA\ break plus EW of different emission lines such as [Ne III] 3869\AA\ or [O II] 3727\AA\ \citep{Stasinska_2006} have all been found to be suitable classifiers.  A combination of all available physical parameters can also be considered, but this approach demands tools and classifiers that can handle such a complex, multi-dimensional problem (e.g., Teimoorinia et al. 2016).  Using such methods, the requirement of obtaining relatively high SNR spectra of several emission lines to classify AGN galaxies and SF galaxies can be circumvented if alternative physical parameters are shown to be efficient classifiers.  Therefore, three important questions regarding the BPT galaxy classification scheme naturally arise: 1) Which physical parameters can reproduce a BPT diagram in the absence of one or more of the four emission lines used for the classification? 2) What is the ranking of importance between such physical parameters? (i.e., are certain combinations of parameters better at classifying than others?)  3) Can spectral information other than emission line flux provide sufficient information to re-construct the diagram?

Galaxies are  complex systems composed of many physical parameters.  Complicated correlations between those parameters can sometimes exist, which require powerful models to be explored.  The precision of predictions  in a problem can be increased, however, if we use more informative parameters and make suitable connections between them. These connections can be found using ANNs, which are models with highly interconnected nodes that make it possible for a deeper exploration of astronomical data sets \citep{Tem_2014, Gonzalez-Martin_2014, Tem_2017}.  ANNs can be configured and trained to find different patterns in a data set, such as distinguishing stars from galaxies \citep{Soumagnac_2015}. Distinguishable patterns can be described by different statistics and then, for example, can be used to rank the relevance of physical parameters in an astronomical problem \citep{Tem_2016}. AGN and SF galaxies are therefore interesting objects from a pattern recognition perspective.

It should be noted that every survey has its own characteristics, which arise out of its specific goals, observational strategies, and data reduction techniques. These differences can put practical limitations on generative models for making useful predictions. So, it is important to note that our approach is, generally,  valid in the context of the SDSS or similarly designed spectroscopic surveys of galaxies. For example, \cite{Trump_2015} show that  different complex selection effects can affect BPT classifications. In other words, the selection effects are applicable only to surveys with a very similar observing strategy. In this respect, \cite{Tem_2017}, by pattern recognition methods, show that their HI mass predictions are limited to a specific parameter space and cannot be generalized to other surveys with different characteristics.  Instead, they present different `control' parameters to distinguish the galaxies that have different patterns (from the training set) and, consequently, show higher uncertainties.  Even a survey such as the Sloan Digital Sky Survey (SDSS) has characteristics that can bias pattern recognition.  For example,  the SDSS fiber covering fraction (due to different galaxy sizes and  wide range of redshifts; $\sim [0.002-0.35]$) can be a problem in capturing `complete' information from galaxies.  This issue can be an important point in presenting a predictive model, but can be minimized when restricting an analysis to galaxies at $z>0.04$ \citep{Kewley_2008}. To have a more homogeneous sample as a training set and reduce the affects of the aforementioned problem, we use a smaller redshift interval in our sample.

Limiting our training set to a small interval of redshift reduces both the effect of distance on observed colors (e.g., g-r) and also the effects of the fiber covering issue. At the same time, we note that the definition and original classification of AGN, composite, and SF galaxies are based on the information obtained within a 3 arc-second-wide fiber footprint. This paper, however, has an informational approach in which the aim is to take the definition (i.e., the defined classes based on 3 arc-second fiber) as a target and demonstrate the potential of physical parameters in re-creating the classes. So, here the aim is not to present a catalog (with different `control' parameters) for higher redshift galaxies, for example, or present predictions for galaxies from different surveys with different strategies.  Instead, to show the robustness of our results, we will use a `test' set comprised of galaxies that are outside the redshift interval chosen for the training set.

In this paper, to answer the questions posed above,  we will rank physical parameters such as the stellar mass, which can be used jointly with \ohb\ to predict the three classes of galaxies in a BPT diagram and show that spectral information,  by itself, can provide better results.  In Sec. \ref{data}, we describe the data used in this paper and in Sec. \ref{method},  we present our method.  The results, discussion, and a short summary are presented in sections \ref{result}, \ref{discussion},  and \ref{conclusion} , respectively. In this paper, we consider a cosmology with $\Omega_{\rm{tot}}$ , $\Omega_{\rm{M}}$, and $\Omega_{\rm{\Lambda}}$ = 1.0, 0.3, and 0.7, respectively, and H$_0$ = 70 km s$^{-1}$ Mpc$^{-1}$.

\section{Data}
\label{data}

Line fluxes (all with $\rm{SNR}>3$,  corrected for Galactic extinction and underlying stellar continuum \citep{Schlegel_1998}), stellar masses \citep{Kauffmann_2003, Salim_2007}, equivalent widths, and other galactic parameters (and the associated  errors) are taken from the Max Planck Institute for Astrophysics (MPA)/Johns Hopkins University (JHU) catalogs of galaxies. The Petrosian photometry  is used in this paper \citep{Blanton_2001, Yasuda_2001}. The ratio of radius containing 90\% and 50\% of the Petrosian flux, the velocity dispersion\footnote{Calculated from Princeton/SDSS spectroscopy, as listed in the MPA/JHU catalog, available at \myurl }, mass to light ratio \cite[in g-band following][]{Kauffmann_2003}, and 4000\AA\ break \citep[i.e., Dn4000; ][]{Balogh_1999} are denoted by R$_{90}$/R$_{50}$,  Vdis, (M/L), and Dn4000, respectively.  In this way and in total, we have a spectroscopic sample with 160922 galaxies at (z $\sim [0.002-0.35]$).

To reduce the redshift problems mentioned in Sec. \ref{introduction}, we limited our training set to galaxies within the redshift interval [0.05-0.15].  Additionally, since the number of AGN and SF galaxies is not balanced in this redshift interval (which is also the case for Seyfert and LINERs), we randomly selected 5000 galaxies from the SF, composite, and AGN classes to create our final training set of 15000 total galaxies.  We call this set the main training set.  To also have a balanced test set including galaxies outside our training set redshift interval, we randomly selected another 15000 galaxies (5000 galaxies from each class) with redshifts z$<0.05$ and z$>0.15$.

\section{Method}
\label{method}
\begin{figure}
\centering
\includegraphics[width=8.5cm,height=5cm,angle=0]{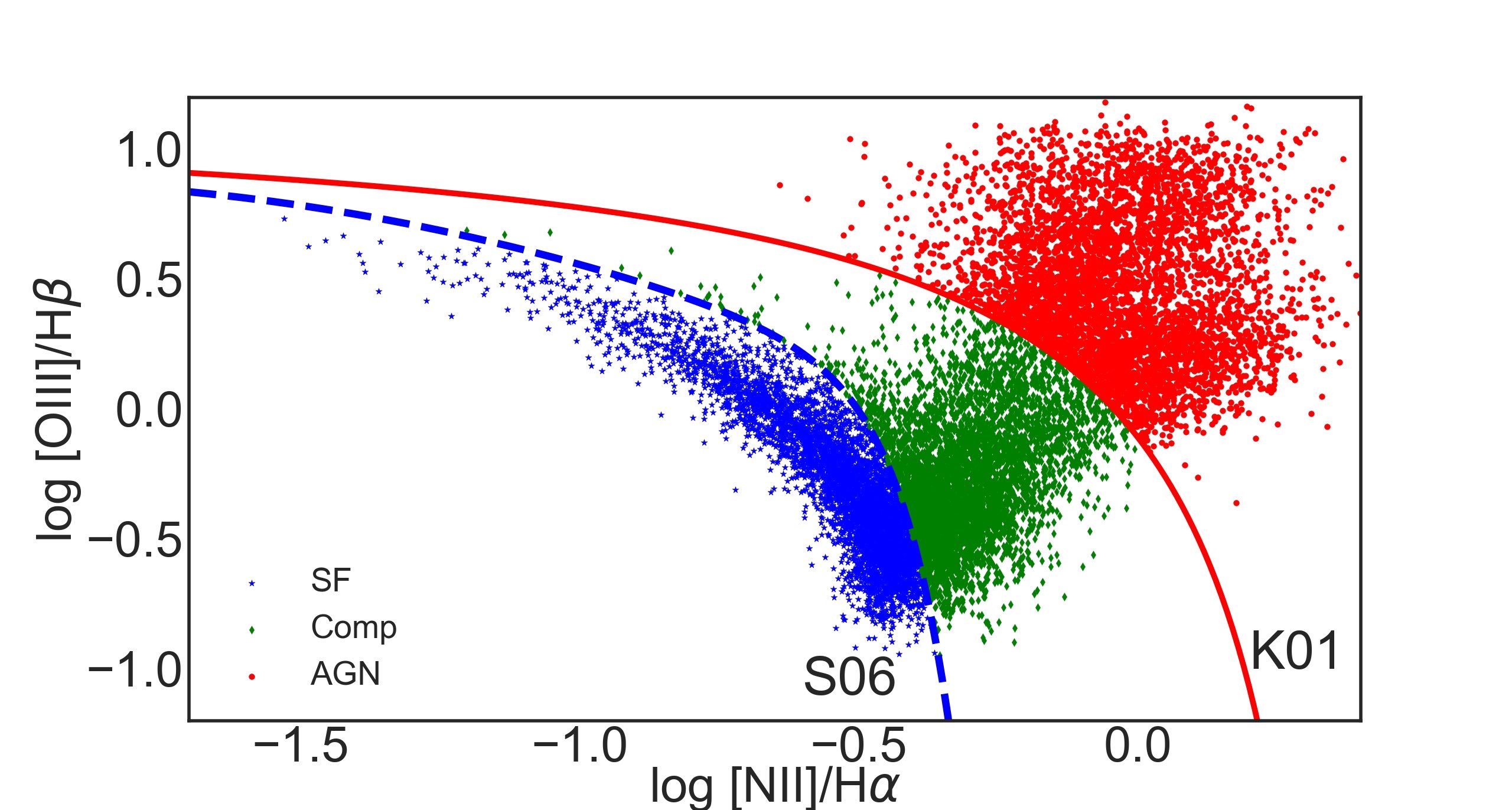}
\includegraphics[width=8.5cm,height=5cm,angle=0]{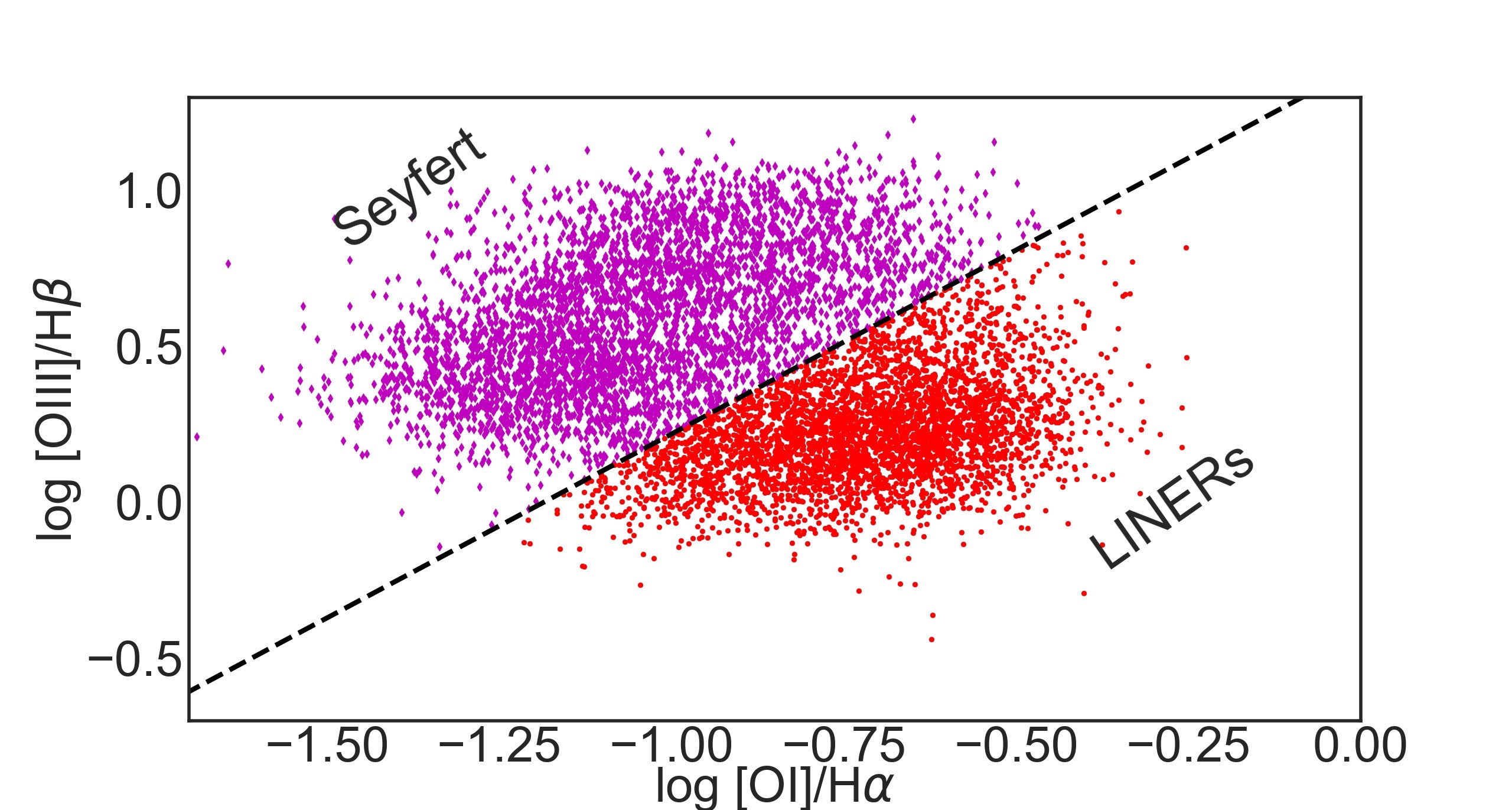}
\caption{The top plot shows 15000 galaxies separated into three different classes based on their position in the  BPT diagram. These galaxies are used as the training set in this work. Galaxies above the red line determined by K01 are classified  as AGN (red points). Galaxies under the blue line determined by S06 represent SF galaxies (blue stars). The green diamonds between the two lines are the composite galaxies.  The bottom plot shows the AGN galaxies after being sub-classified into Seyferts or LINERs. }
\label{fig-bpt-tr}
\end{figure}

The top panel of Figure \ref{fig-bpt-tr}  shows a BPT diagram that contains 15000 galaxies, i.e., the main training set (shown in three different colors based on their BPT classification) described in Sec. \ref{data}.  The blue dashed line in Figure 1 separates star-forming galaxies (the blue `star' points) from the rest of the galaxies. This line is drawn based on theoretical estimates that are presented  by S06. The red solid line is proposed by K01 as the  curve that  separates AGN galaxies (the red `circle' points) from other galaxies. The green `diamond' points that lie between the S06 and K01 lines are the composite galaxies.  In addition to the flux information of the four emission lines, we also have additional information such as stellar mass and photometric colors for each galaxy in this plot.  In the bottom panel of Figure \ref{fig-bpt-tr},  the AGN galaxies (from the top plot) are classified into two sub-classes according to the method introduced by K06. In this method, the black dashed line separates LINERs and Seyfert galaxies as shown in the plot. 

As mentioned in Section 2, the main training and test sets are balanced sets in which the number of galaxies is the same in each class.  We further divide the main training set into a `new' training set (70\% randomly selected from the main training set)  and a `validation' set (30\%), which is different from the test set introduced in Sec \ref{data}. To avoid over-fitting, we use the early stopping method in which the training procedure stops if the `new' training and validation set begin to show different performance. Although this step ensures that the predictive power of our final models is similar between the training and validation sets, the ranking of parameter sets in this paper are obtained using the validation set.

In a pattern recognition problem, which is a supervised method, we need a set of physical parameters as the input to a network  and the correspondence labels as the target data. Then the  connections between the input parameters and targets can be made by the network. After training the network (i.e., fixing the network's internal parameters), the class (or label) of a new object can be predicted based on its input information and probability theories \citep{Bishop_2007}. The highly interconnected  nature of ANN's models allow us to make all  possible  connections between input parameters  and  also between the input and the target in different layers. We use a three-layer network in this paper.  In this way, for example,  all the possible connections between a set of input  parameters  and the three classes of galaxies (shown in the top plot of Figure \ref{fig-bpt-tr})  can be made. We will repeat this procedure for a two-class case (i.e., for the bottom plot of Figure \ref{fig-bpt-tr}). First, we will present several examples of the three-class classification. In each example, we use different input data while keeping the three target classes the same to show the effect different inputs have upon the performance of the trained network.

In the simplest case, we first use only \ohb\ and connect this parameter to the three classes via a network. This case is a trivial example which is not more complicated than an averaging process. It does provide, however, a convenient example to explain the methodology used for the more complex cases involving additional input parameters that will be discussed in subsequent sections. For making the target data, the classes of SF, composite, and AGN are introduced as three labels (1, 0, 0), (0, 1, 0), and (0, 0, 1), respectively.  After training the network,  the output result for a galaxy will be a set of three probabilities.  As an example, a set of probabilities such as  (0.85, 0.1, 0.05) shows that the galaxy under study most likely belongs to the SF class. If the predicted result (label with the highest probability) is different from the true label, then we have a confusion. All possible confusions as well as the correct predictions, for all the 15000 galaxies,  can be arranged in a matrix so called a confusion matrix.  To obtain statistical results and also avoid over-fitting problems, we use the methods and techniques described in \cite{Tem_2012}, \cite{Tem_2014} and \cite{Ellison_2016b}.

\begin{figure}
\centering
\includegraphics[width=9cm,height=7cm,angle=0]{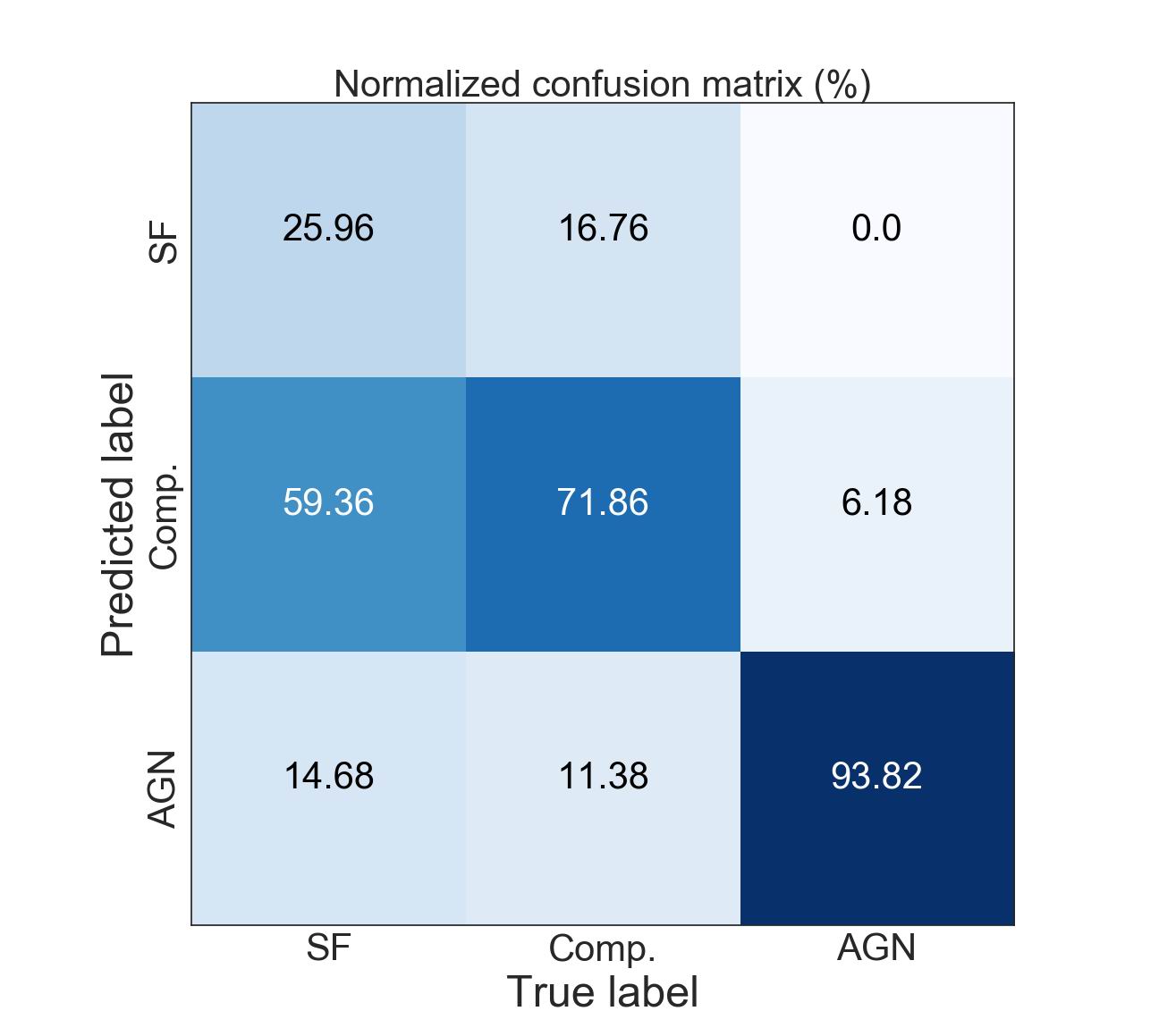}
\includegraphics[width=9cm,height=5.5cm,angle=0]{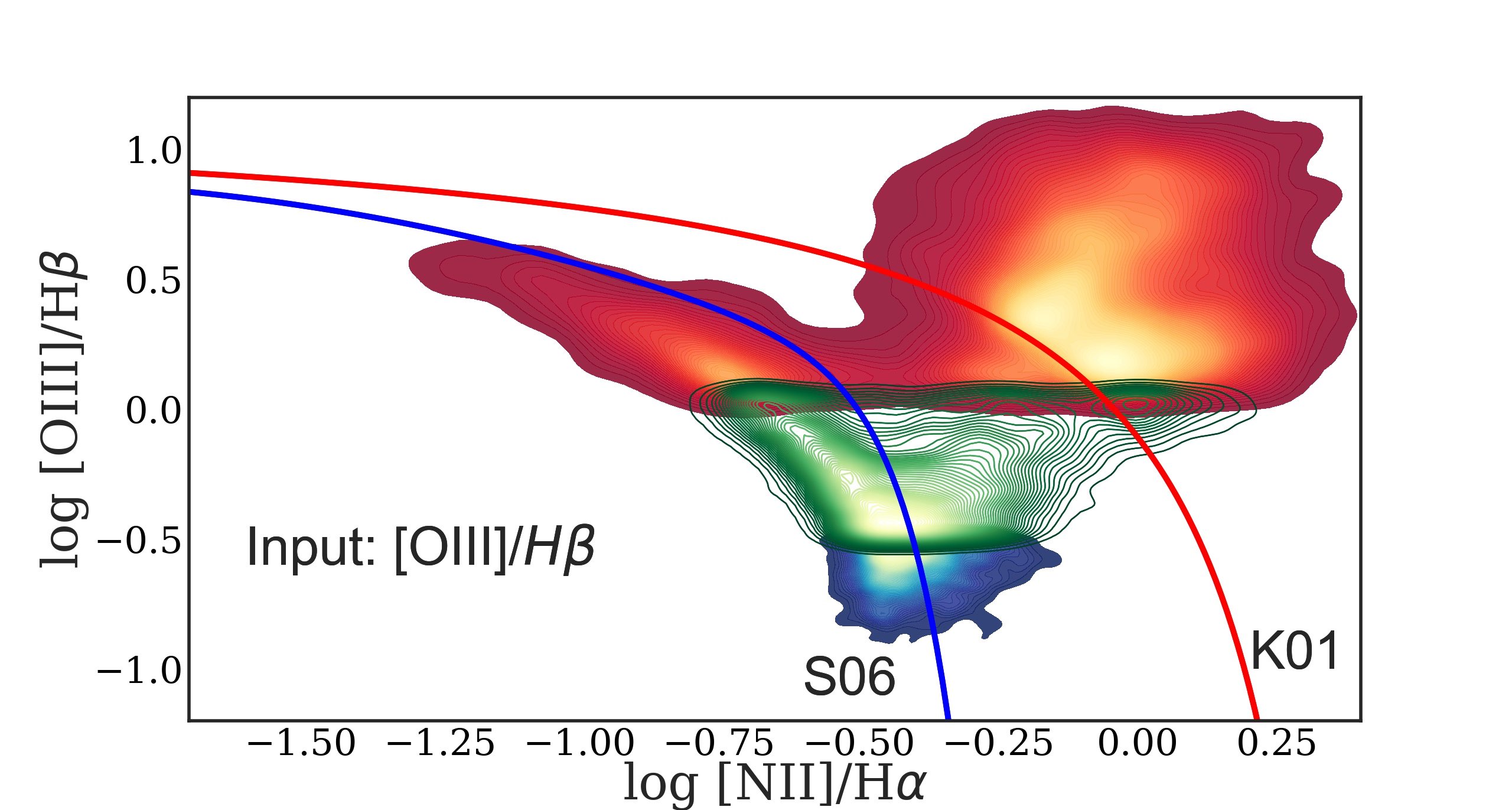}
\caption{The top plot shows a confusion matrix when we use only \ohb\ as the input data in the pattern recognition problem. The most populated areas are shown with darker colour.  AGN  galaxies can be recognized with more that 93 percent probability. The SF group is the least predictable  group. The confusion between the SF and AGN groups is $\sim$14.7\%. The highest confusion involves over 59\% of the SF galaxies that have been falsely predicted to be composite galaxies.  The bottom plot shows the predicted results displayed on a BPT diagram. Here, the predicted AGN, composite, and SF galaxies  are shown as red, green and blue colours, respectively.  The brighter areas show a higher density of points. }
\label{fig-cm-xoh}
\end{figure}

The top plot of Figure \ref{fig-cm-xoh} shows a confusion matrix for the case in which the input data contains only \ohb.  We have 5000 galaxies in each class but for easier comparison, the result in the confusion matrix is described in terms of percentages. The vertical square boxes are the true labels (i.e., the labels taken from the top plot of Figure \ref{fig-bpt-tr}) in which the total percentage adds up to one. The horizontal boxes show the network's predictions (i.e., the statistical results). In the matrix, more populated areas are shown with darker colour. A perfect result can happen when all diagonal squares  show a number of 100\%.  A perfect result can be easily  obtained if we use both \nha\ and \ohb\ as input data. The confusion matrices are examples of loss functions in which the aim is to minimize the percentage of the off-diagonal squares (or maximize the diagonal percentages).  The BPT diagram displayed in the bottom plot of Figure \ref{fig-cm-xoh} shows the predicted AGN, composite, and SF galaxies as red, green, and blue contours, respectively.  This color scheme is used in similar plots throughout the paper to denote each predicted class.  

As can be seen from the confusion matrix in Figure \ref{fig-cm-xoh}, 14.68\% of SF galaxies are falsely classified as AGN galaxies. This is an example of a confusion. In the bottom plot of the Figure, this percentage is related to the red area under the blue S06 line.  As another example,  6.18\%  of the AGN  galaxies are predicted to be composites. This percentage is the green area above the red K01 line.  And finally,  59.36\% of star-forming galaxies are falsely predicted to be composite galaxies (which is a large confusion).  Thus, it can be seen that with \ohb\ as input data, AGN galaxies are the best-predicted class and SF galaxies are the worst-predicted class.

To compare multiple classifications that utilize different sets of input parameters, we need to consider the false positives and true negatives. On the other hand, a single number can be more useful for direct comparisons of multiple networks that have been trained using different sets of input parameters. We could obtain an  average as a  single number from the diagonal members from confusion matrices. In doing so, however, we would ignore the false positives which have a direct effect on the classifications. 
 
\begin{figure}
\centering
\includegraphics[width=8.5cm,height=7cm,angle=0]{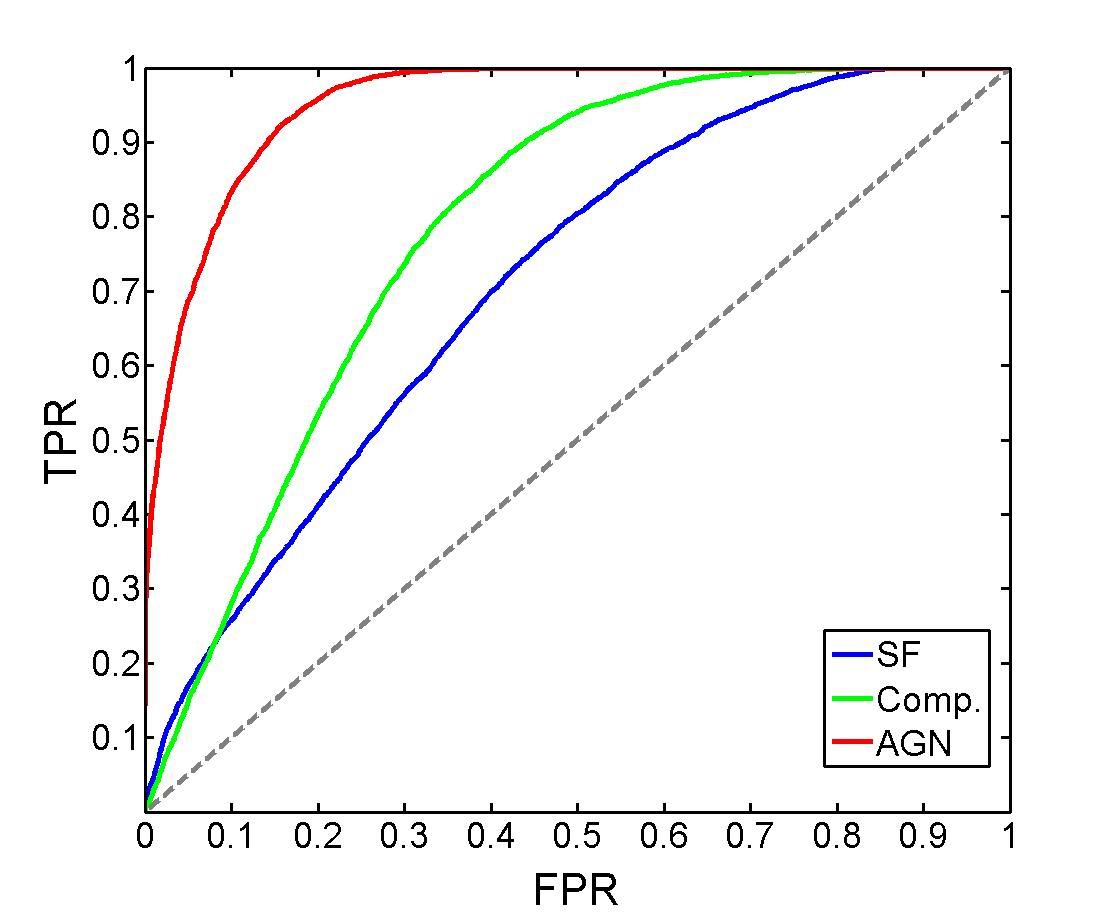}
\caption{ROC plot for the three-class classification presented in Figure 2. The horizontal and vertical axes show False Positive Rate (FPR) and True Positive Rate (TPR), respectively.} The gray dashed line shows a random classification with AUC=0.5. The highest and lowest values are related to AGN and SF galaxies, respectively, showing that the trained network is better at predicting AGN galaxies than SF galaxies.
\label{fig-auc-oh}
\end{figure}

While confusion matrices are very useful tools to visualize and  check a result, there are other useful methods by which we can obtain a single number for convenient comparisons.   In this respect, we use Receiver Operating Characteristic (ROC) plots, which are used to estimate the statistical results of a classification (see \cite{Tem_2016} and reference therein). Briefly, the area under the curve (AUC) of the plots  can be a number between  0.5 and 1, which represent  random and perfect classifications, respectively.  The average AUC for the classification presented in Figure \ref{fig-cm-xoh}  is $\sim0.82$. This number can be compared with the results from networks trained using different inputs. In the next step, we will change the input.  An example of a ROC plot is shown in Figure \ref{fig-auc-oh}. The horizontal and vertical axes show False Positive Rate (FPR) and True Positive Rate (TPR), respectively.  The gray dashed line shows a random classification with AUC=0.5.  As can be seen, AGN have the highest AUC of the three classes and thus the highest rate of correct classifications.  Meanwhile, SF galaxies have the lowest AUC and thus the lowest rate of correct classifications.

\begin{figure}
\centering
\includegraphics[width=9.8cm,height=7.7cm,angle=0]{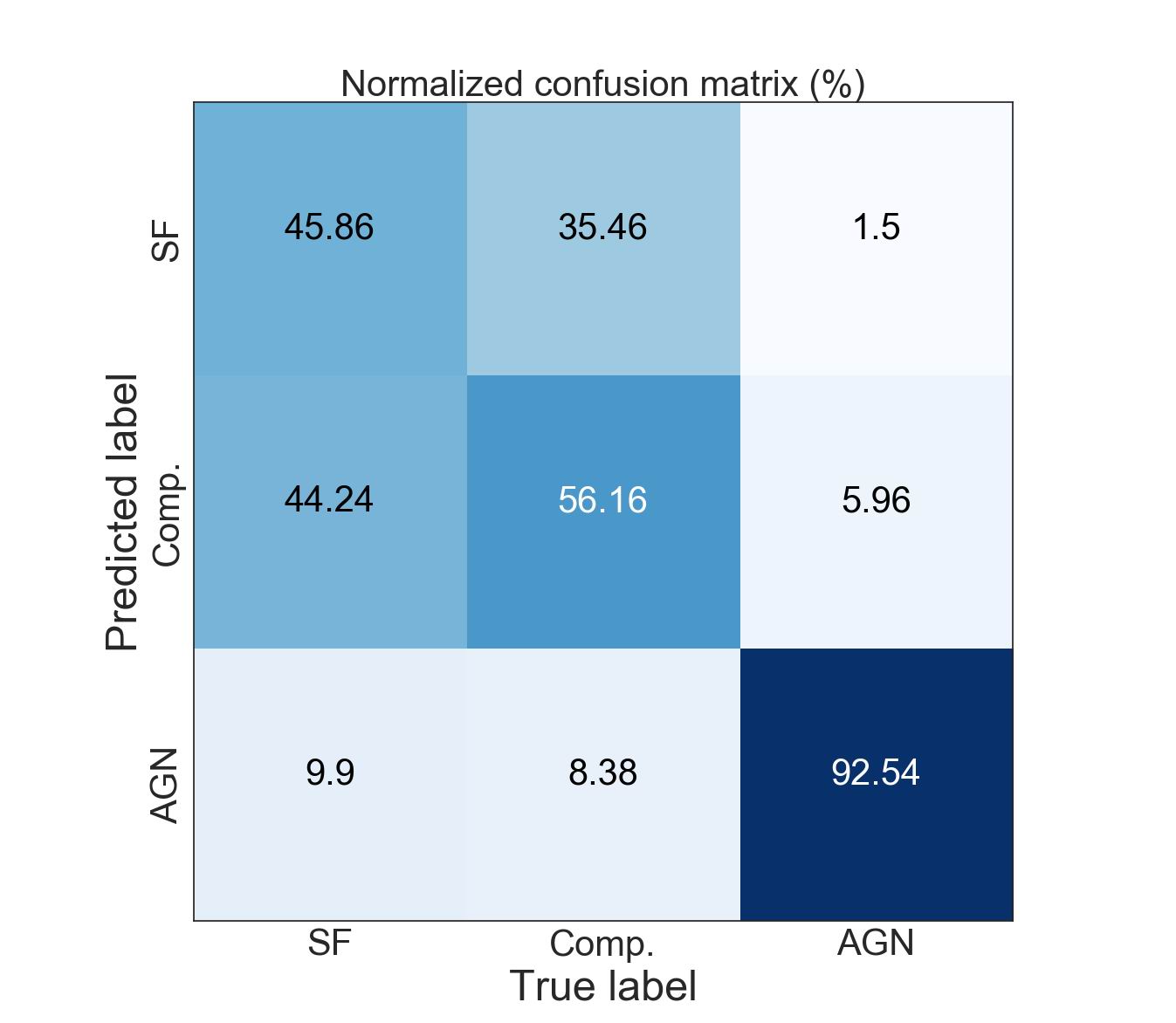}
\includegraphics[width=9cm,height=5.5cm,angle=0]{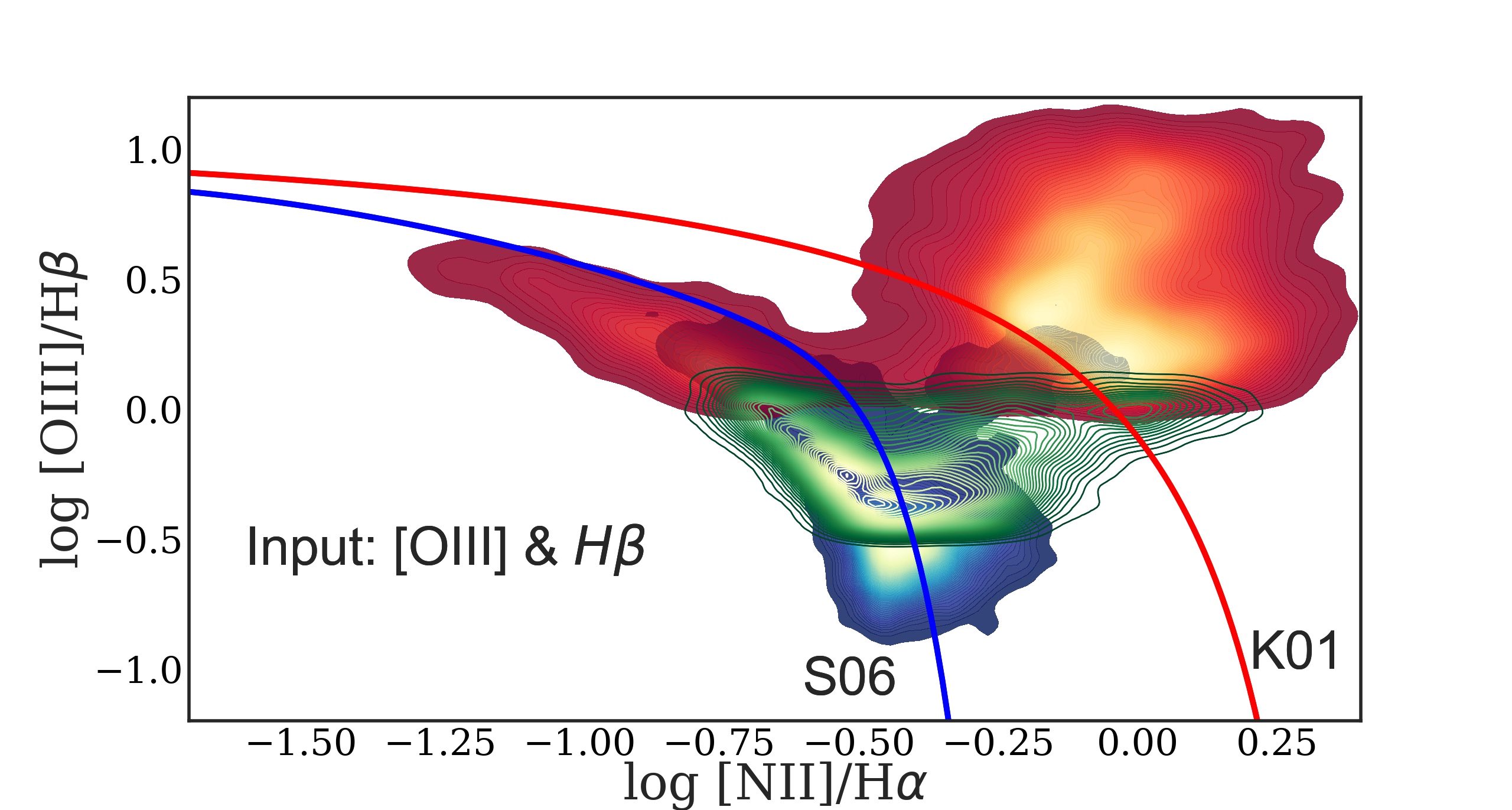}
\caption{This figure is the same as Figure \ref{fig-cm-xoh} when we use the flux of [OIII] and H$\beta$ as two separate parameters instead of their ratio. The SF-SF box shows a better result with respect to Figure \ref{fig-cm-xoh}.}
\label{fig-cm-xoh2}
\end{figure}

Figure \ref{fig-cm-xoh2} is the same as Figure \ref{fig-cm-xoh} when we instead use the flux of the [OIII] and H$\beta$ emission lines as two separate input parameters to the network. The average AUC for this classification is $\sim 0.84$ which shows a small improvement with respect to the input as a ratio shown in Figure \ref{fig-cm-xoh}.  The main issue in  Figure \ref{fig-cm-xoh} is related to the SF galaxies. Here, SF galaxies are more correctly classified.  This result can also be seen in the bottom plot of Figure \ref{fig-cm-xoh2}. There is, however, not a big difference between Figures \ref{fig-cm-xoh2} and \ref{fig-cm-xoh} and we can still see large confusions.  In both Figures, most of the SF galaxies are falsely predicted as composite or AGN galaxies. Conversely, more than 90\% of AGN galaxies are correctly classified in both Figures.  Although this result appears to be favorable, there is a considerable amount of false positive galaxies in the SF class. To overcome this issue, we should add a parameter that can better separate AGN and SF galaxies. Ju11  use stellar mass and \ohb\ to obtain a more precise  classification. We find similar improvement when adding information such as mass as input to our network, which decreases the SF galaxy false positives rate and does not change the AGN-AGN predictions considerably.  In other  words, and as can be seen from the confusion matrices,  using \ohb\ as input guarantees the AGN-AGN  prediction is always more than 90\%. This number can roughly be predicted   when we  pay attention to  the distribution of galaxies along the \ohb\  axis in a typical BPT diagram.

\begin{figure}
\centering
\includegraphics[width=9cm,height=8cm,angle=0]{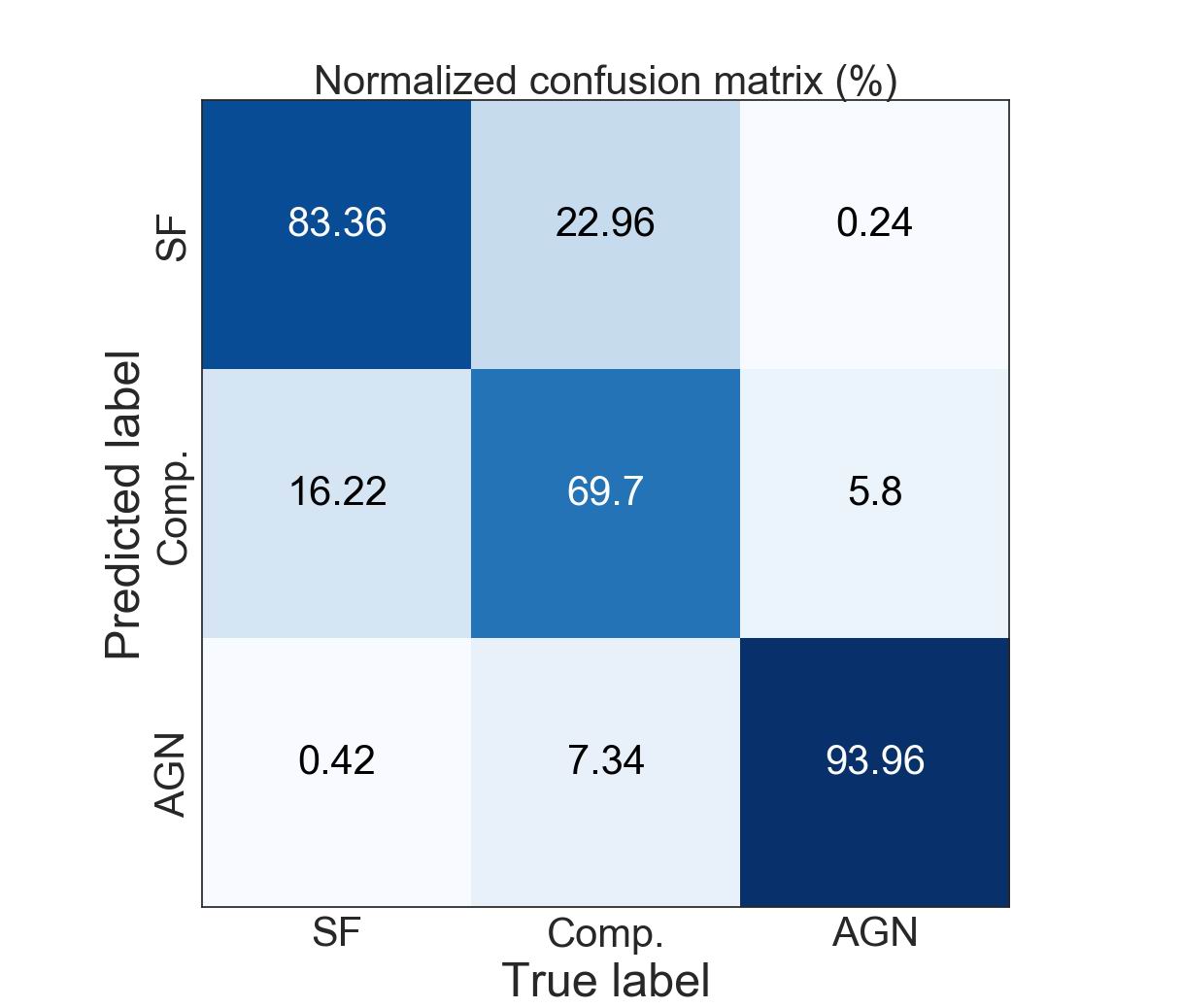}
\includegraphics[width=8.5cm,height=5cm,angle=0]{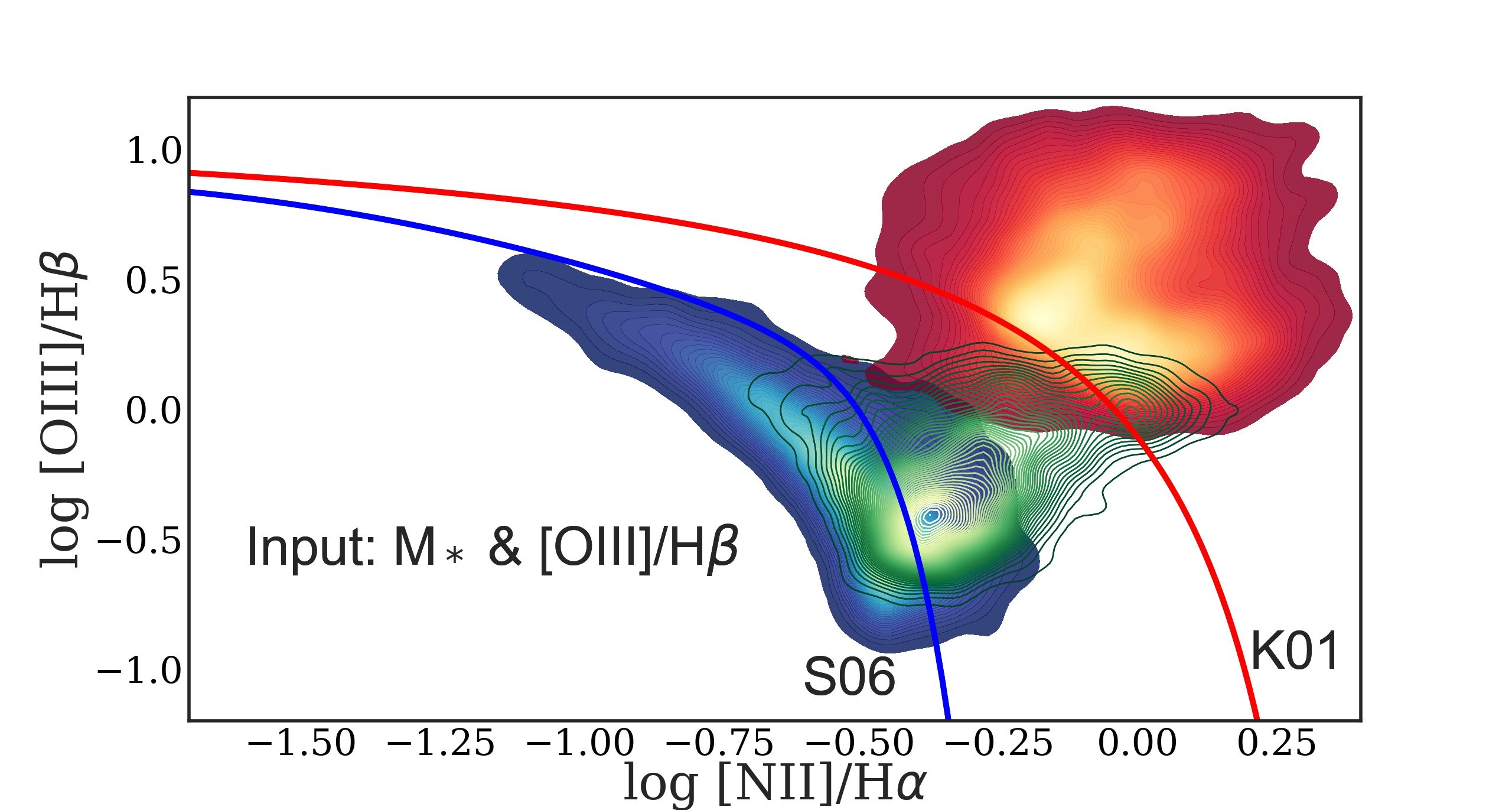}
\includegraphics[width=8.5cm,height=5cm,angle=0]{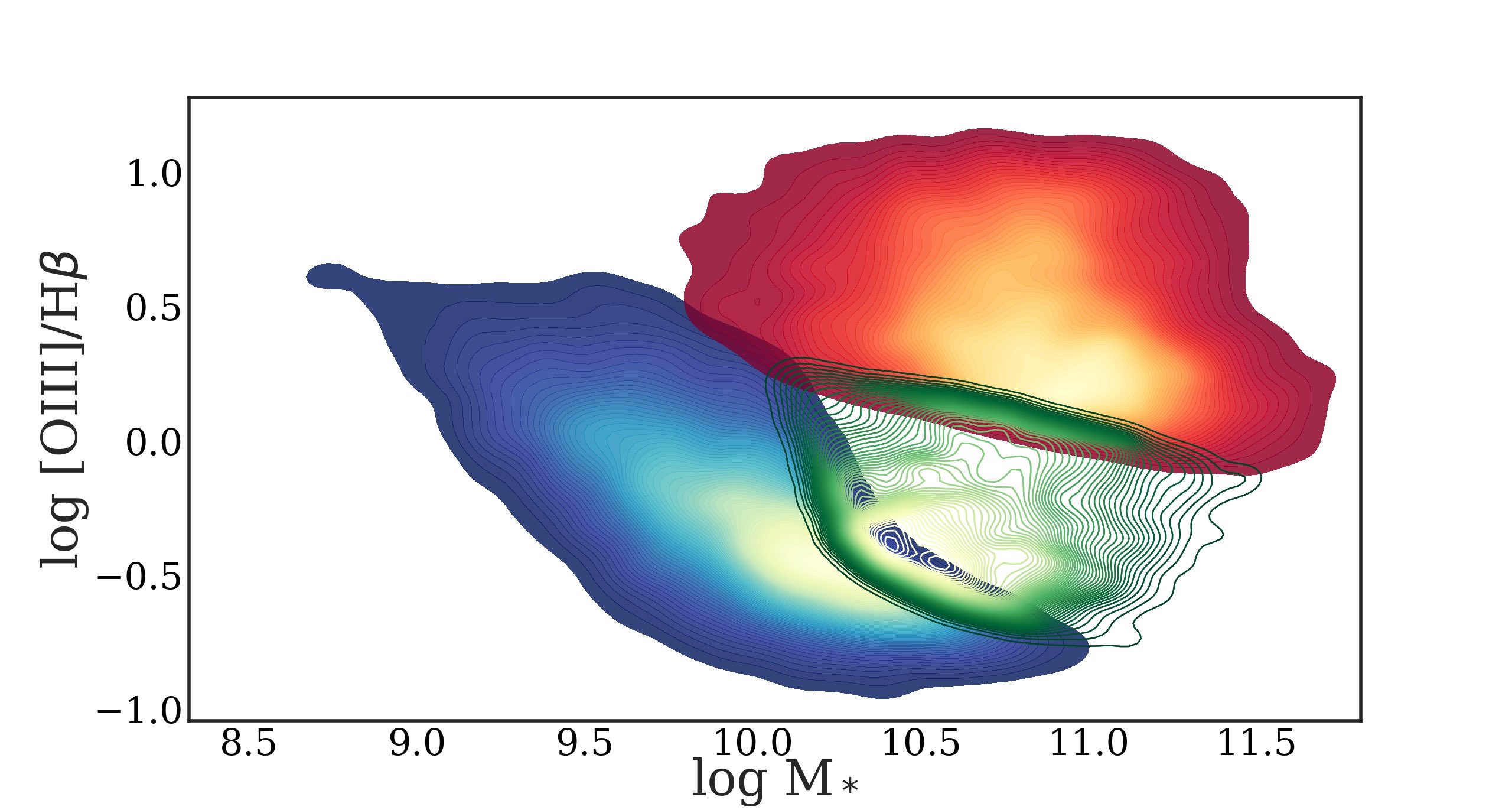}
\caption{The two top plots are the same as Figures \ref{fig-cm-xoh} and \ref{fig-cm-xoh2} when the input parameters are stellar mass and \ohb. A significant reduction is seen in the false-positive rates of the SF and composite galaxies. The bottom plot shows the distribution of \ohb\ versus stellar mass. There is small overlap (and thus confusion) between the predicted AGN galaxies (red area) and SF galaxies (blue area).}
\label{fig-cm-p2xoh}
\end{figure}

Following Ju11, we add stellar mass to \ohb\ as input data, re-train the networks, and show the average results in Figure \ref{fig-cm-p2xoh}.   A significant improvement in the confusion matrix can be seen. Significant change, however, is not seen for the AGN-AGN boxes.   There is a considerable reduction in false negatives in the new confusion matrix, which is shown in the two top plots of the Figure. In the middle plot, we show the BPT diagram taken from the new result. As can be seen, there is little confusion between SF galaxies and AGN galaxies. The confusion related to AGN-Comp and SF-Comp, however,  is not small.  In fact, there is a continuous  transition  from SF galaxies to composite  galaxies (and also from the composites to AGN)  causing higher confusion between them.  In the bottom plot of  Figure \ref{fig-cm-p2xoh}, we show  the distribution of \ohb\ versus stellar mass (the same predictions in the three colors). As can be seen,   the overlap  between the red and blue areas is very small.  After adding mass to the input training set for the network, the average AUC becomes $\sim0.944$, which is a significant improvement over the networks trained with only emission line fluxes as input.

\begin{figure}
\centering
\includegraphics[width=9cm,height=8cm,angle=0]{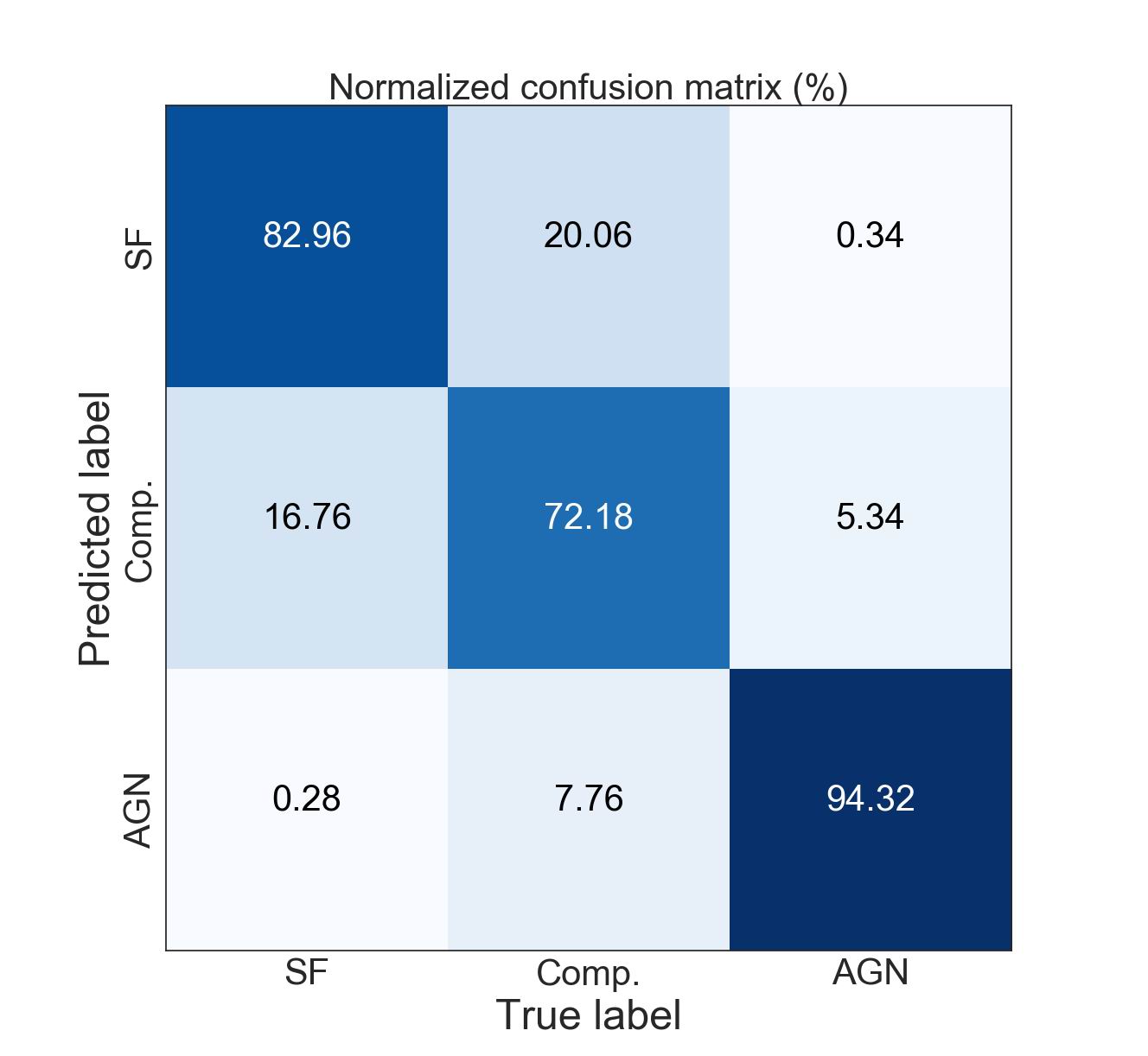}
\includegraphics[width=8.5cm,height=5cm,angle=0]{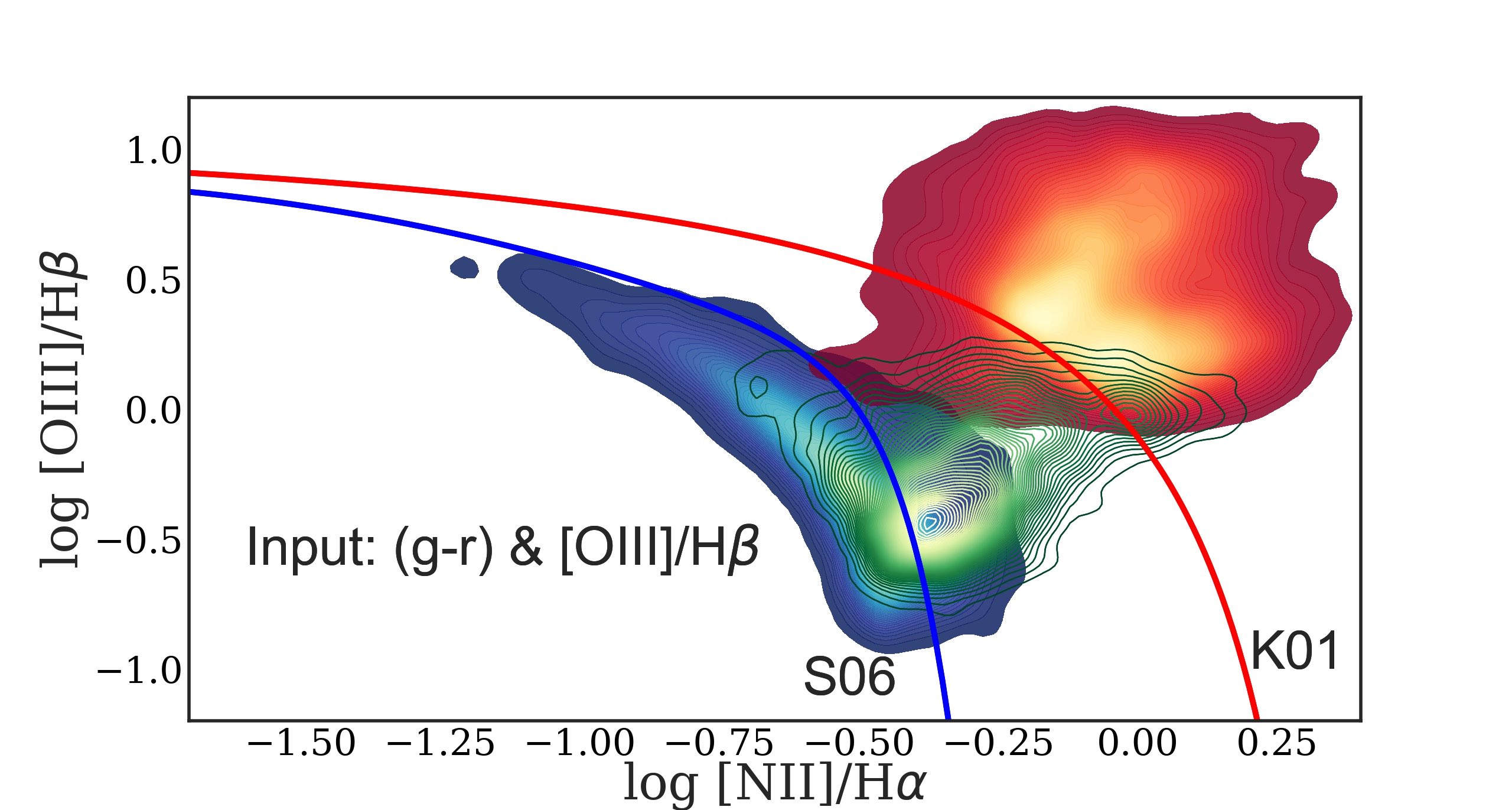}
\includegraphics[width=8.5cm,height=5cm,angle=0]{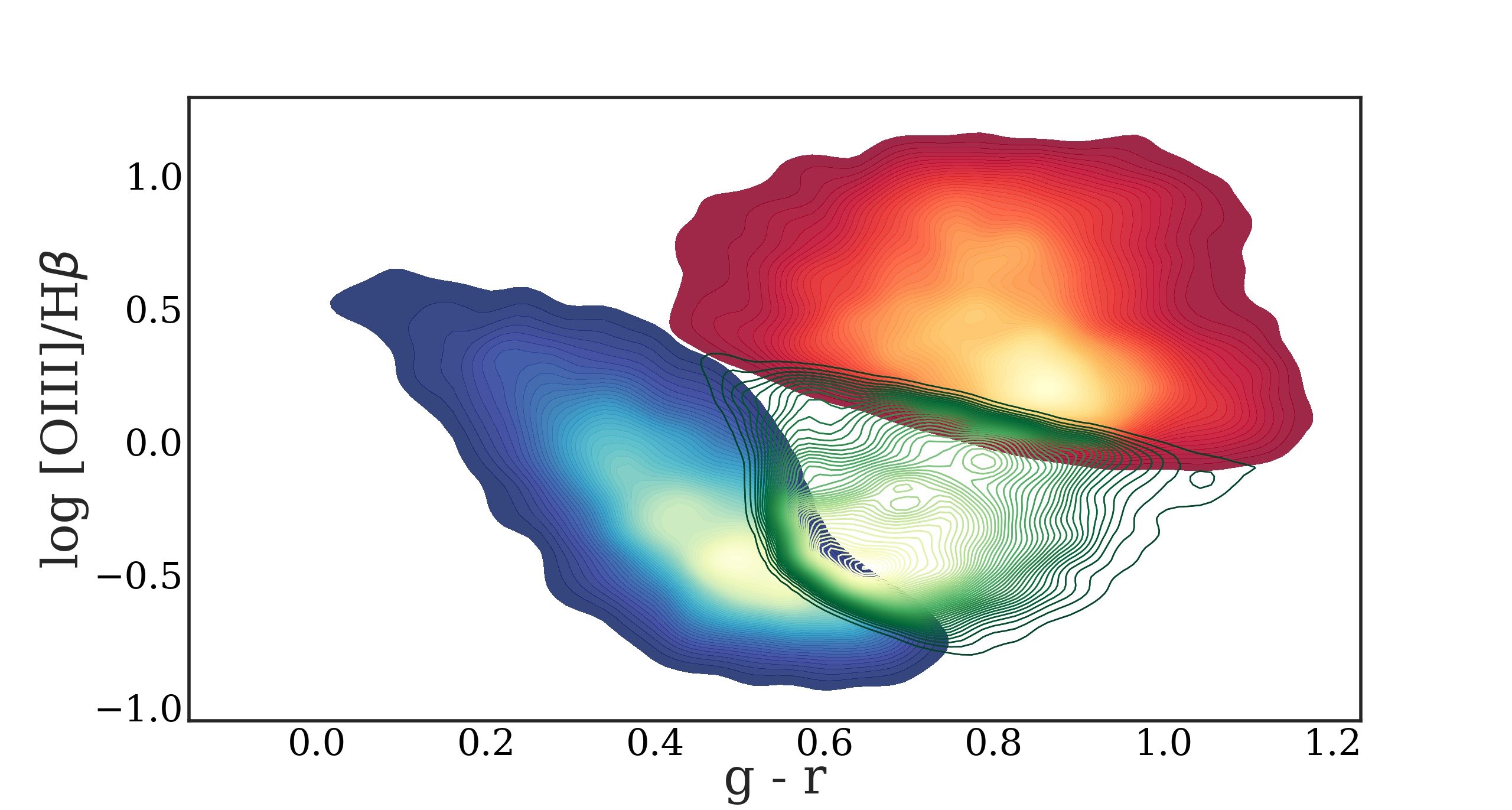}
\caption{This figure is the the same as Figure \ref{fig-cm-p2xoh} when the input contains g-r and \ohb. }
\label{fig-cm-p5xoh}
\end{figure}

The stellar mass is not the only parameter that can significantly improve the results when used jointly with \ohb. For example, in Figure \ref{fig-cm-p5xoh}, instead of stellar mass, we use $\rm{g-r}$ colour. In this case, a significant improvement, similar to that seen when adding mass as input, is observed. For example, the AUC associated with this Figure is also $\sim0.944$.  In the next section, we will present the results obtained after adding additional physical parameters as input data.

\section{Results}
\label{result}

\begin{figure}
\centering
\includegraphics[width=9cm,height=5cm,angle=0]{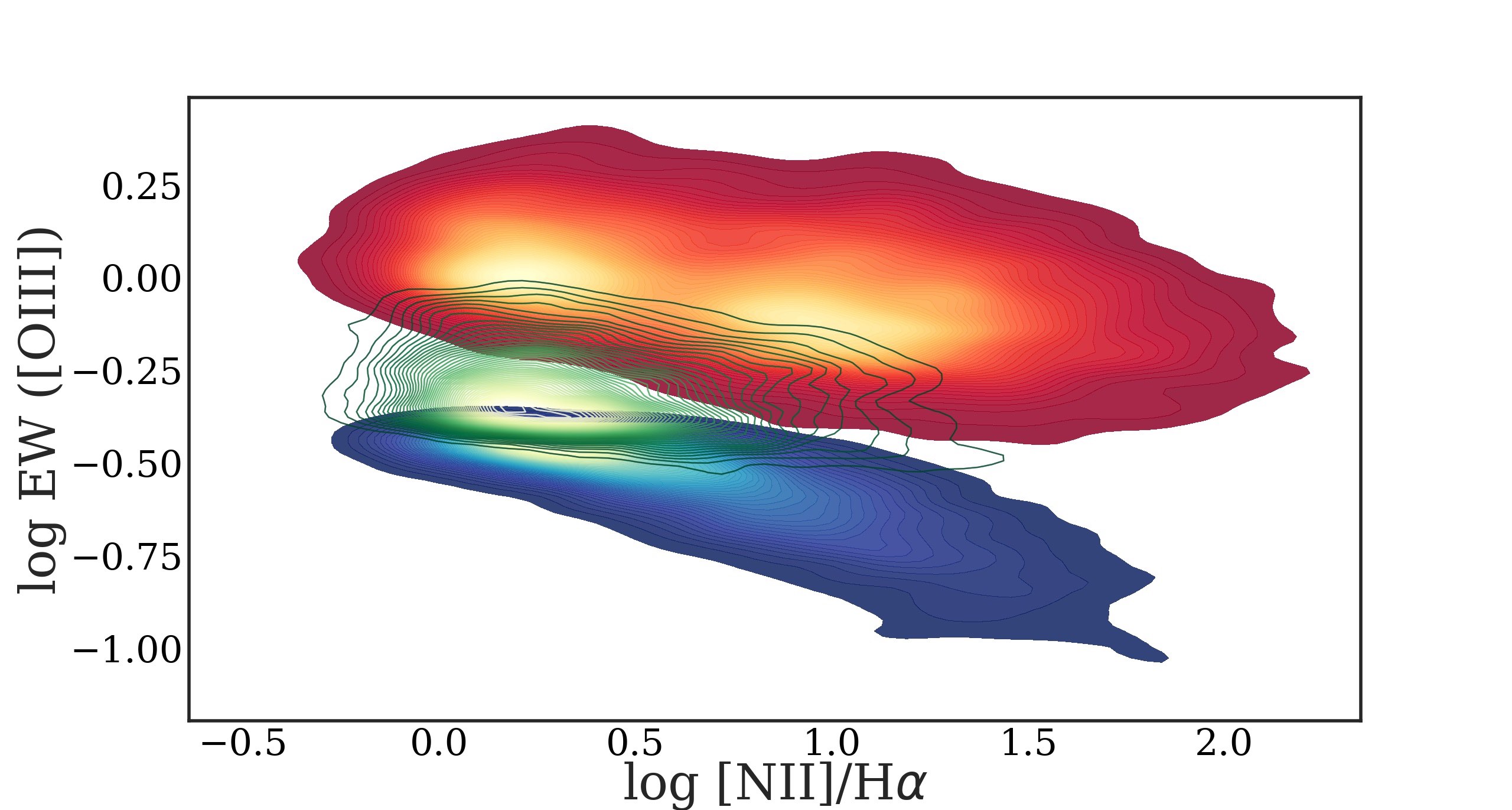}
\includegraphics[width=9cm,height=5cm,angle=0]{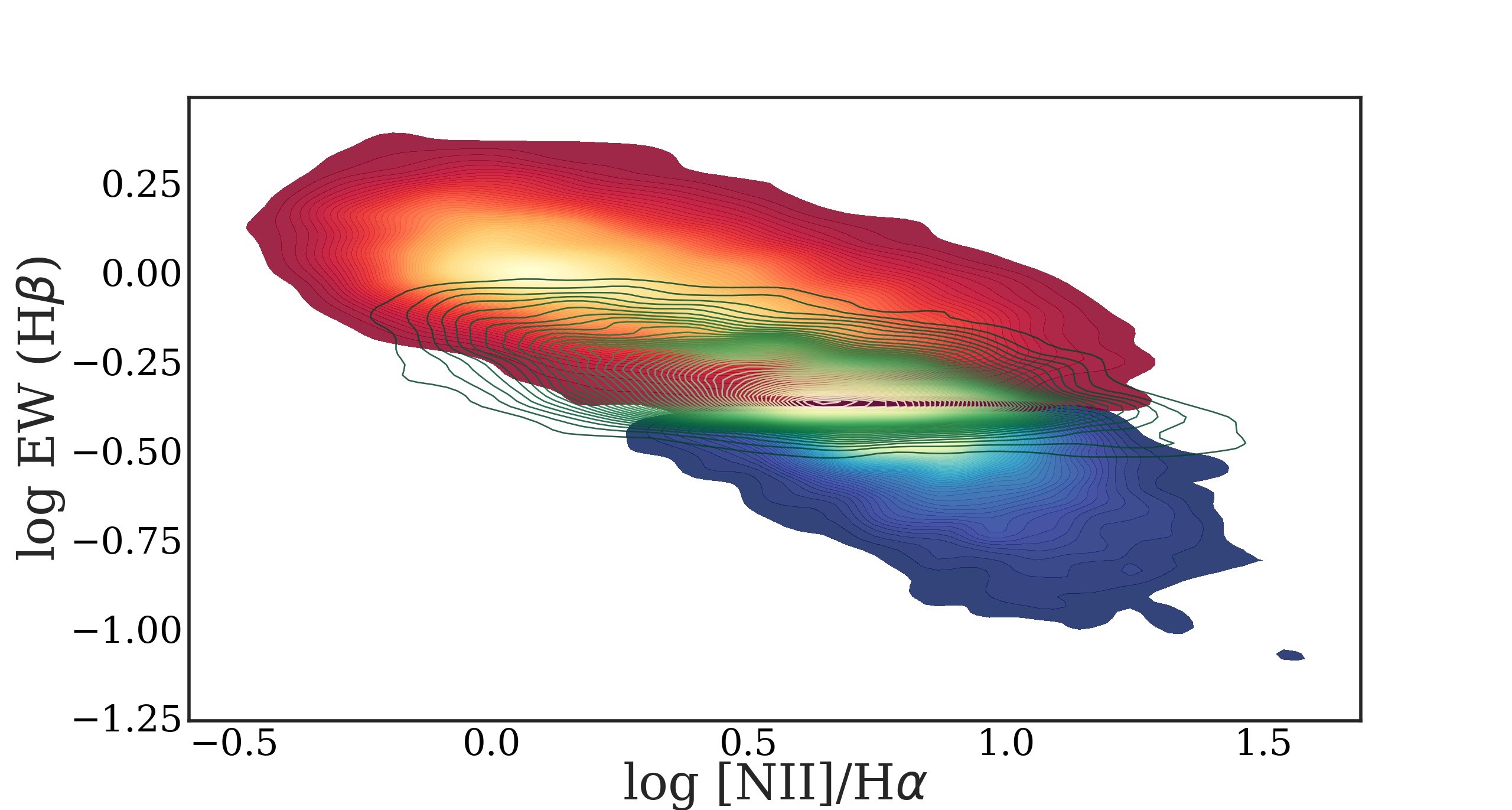}
\caption{The top and bottom plots are the observed EW of [OIII] and \ha\ vs. \nha, respectively. The correlations in both plots can be captured by ANN and can help to improve the confusion matrices.  The color scale is the same as Figure \ref{fig-cm-xoh}.}
\label{fig-ew}
\end{figure}

\begin{figure}
\centering
\includegraphics[width=8cm,height=6.5cm,angle=0]{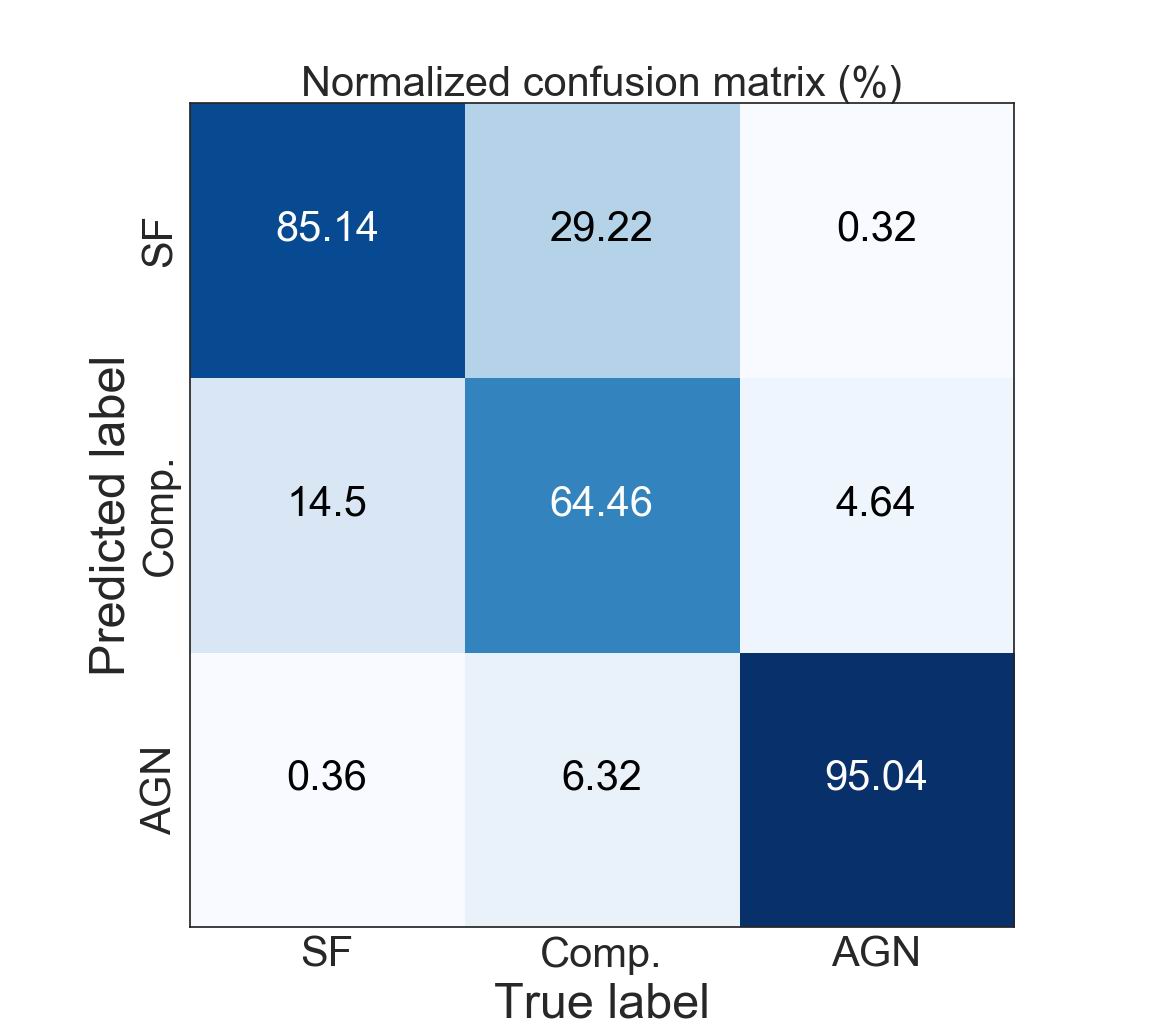}
\includegraphics[width=8cm,height=4.5cm,angle=0]{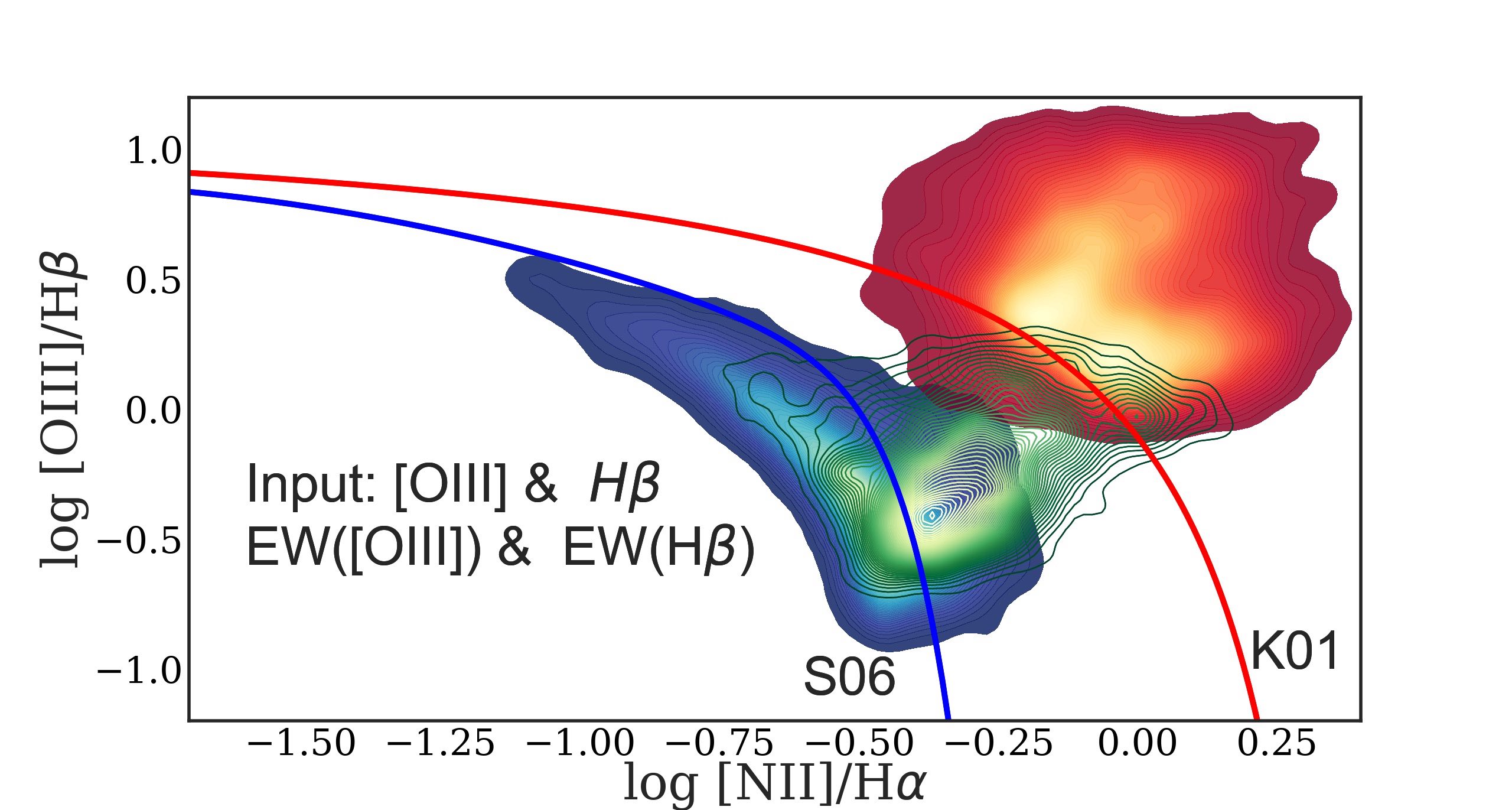}
\includegraphics[width=8cm,height=4.5cm,angle=0]{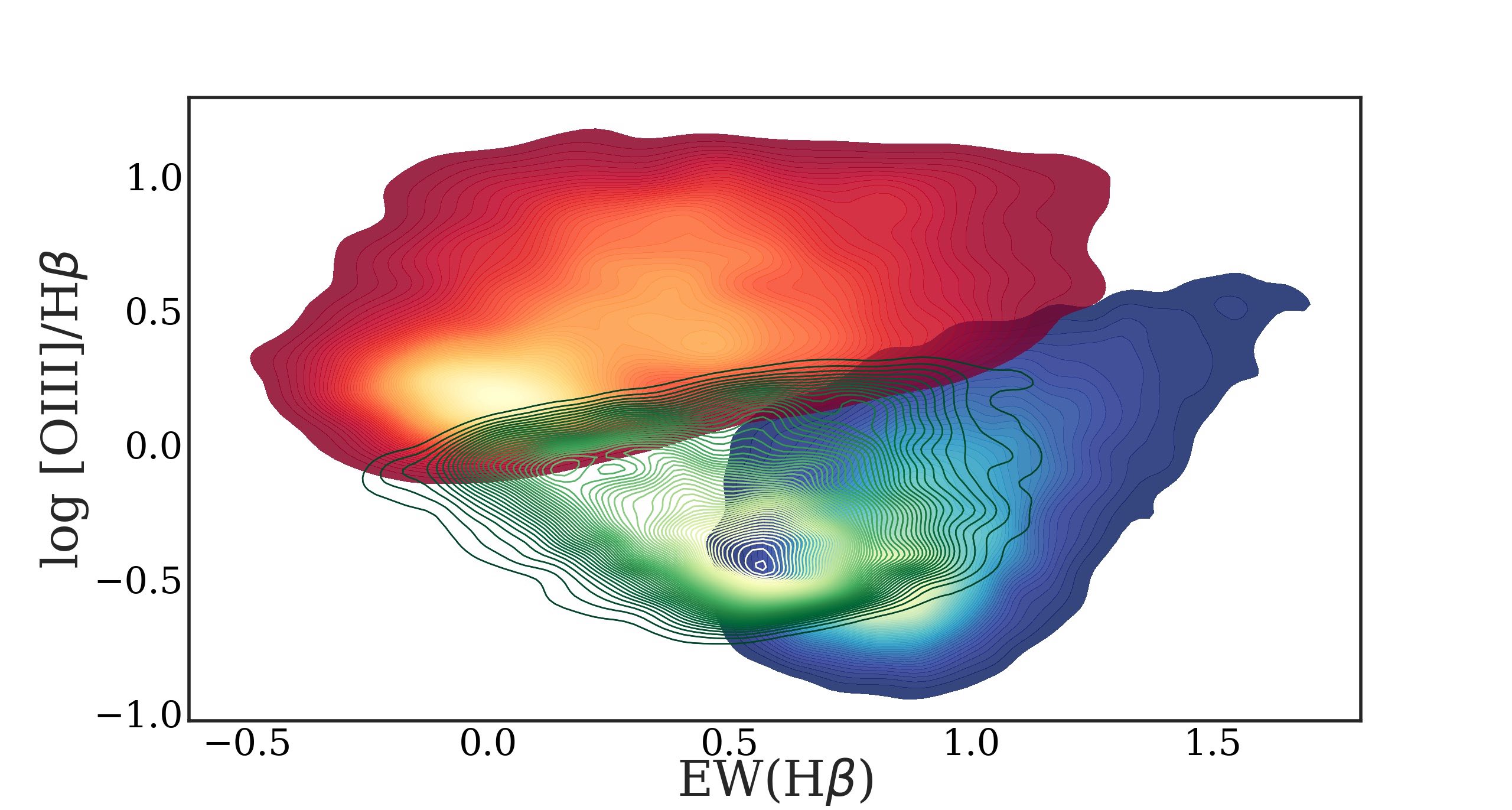}
\includegraphics[width=8cm,height=4.5cm,angle=0]{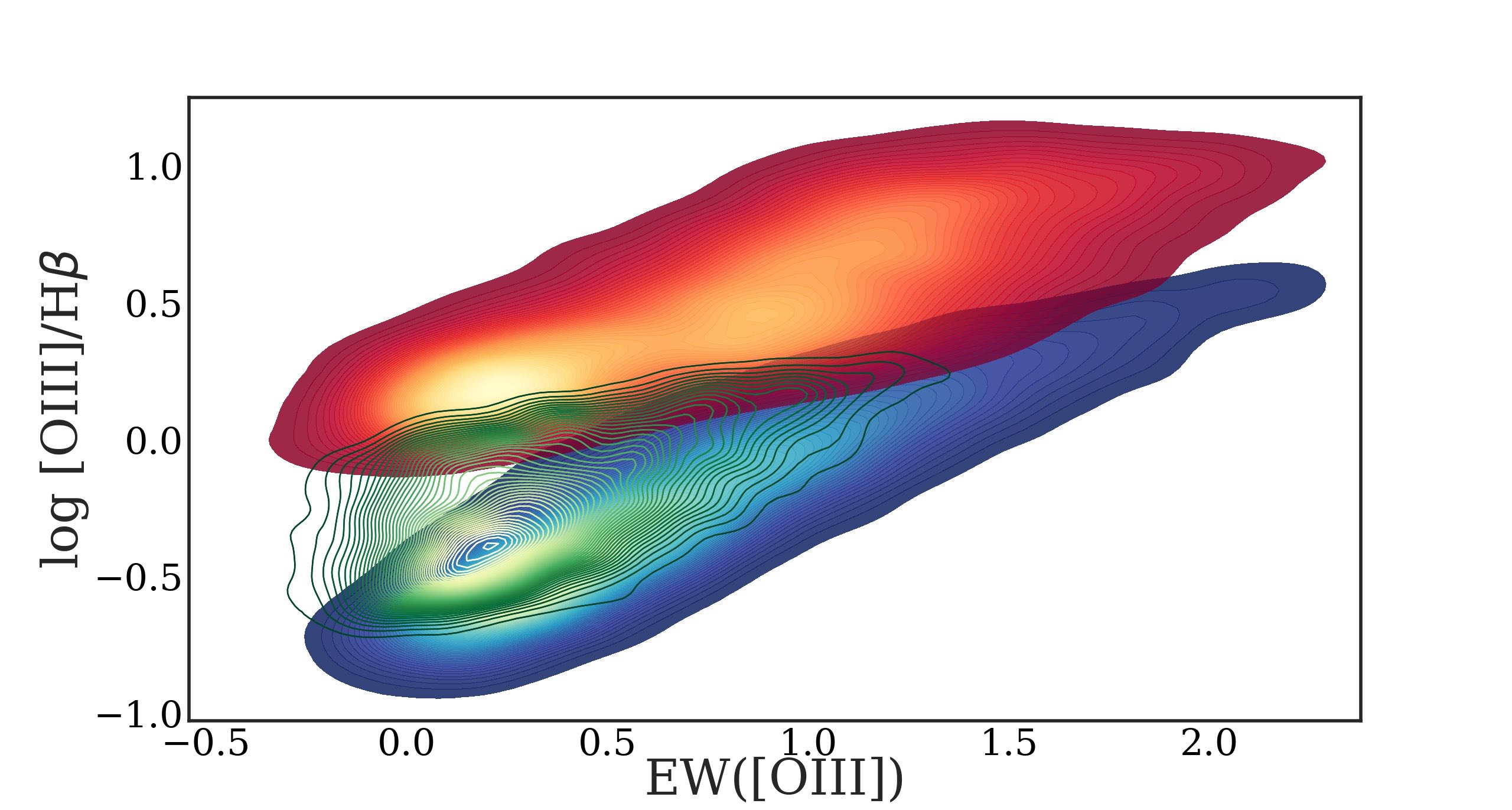}
\caption{The two top plots are the same as Figure \ref{fig-cm-p2xoh} when the input contains the two emission lines [OIII] and \ohb\ as well as their equivalent widths. The two bottom plots show the distribution of \ohb\ versus EW of H$\beta$ and [OIII]. The color scale is the same as Figure \ref{fig-cm-xoh}.}
\label{fig-cm-xoh4}
\end{figure}

The main goal of this paper is to show that spectral information, in the absence of emission lines [NII] and \ha, is capable of predicting the class of AGN galaxies and SF galaxies in the SDSS.  Here, we  will use only the information of the two emission lines H$\beta$ and [OIII]. This time, however, we add the equivalent width of these lines to the input data. The input, therefore, contains four parameters from the spectra. In the top and bottom of Figure \ref{fig-ew},  we show the EWs of [OIII] and \hb\ vs. \nha\ , respectively. These are informative plots that show  distinguishable  patterns from a classification point of view. The correlations between the two EWs and \nha\ can be captured by networks and, as a result, can help to improve the confusion matrices.   The result of the trained network using these additional spectroscopic inputs is shown in Figure \ref{fig-cm-xoh4}. The top two plots of this Figure show the performance of this network is similar to that presented in Figure \ref{fig-cm-p2xoh}. In other words, the EWs have relevant information, similar to mass and g-r colour, which helps reduce the confusion between the three classes.   The two bottom plots of Figure \ref{fig-cm-xoh4} show the distribution of \ohb\ versus  EW of H$\beta$ and [OIII], respectively.   These plots show that AGN galaxies are sufficiently distinguishable from SF galaxies when using the four displayed pieces of spectroscopic information. These distributions, along with the distributions of the [OIII] and H$\beta$ fluxes, are used to predict a probability that a given galaxy belongs to each of the three classes.

\begin{figure}
\centering
\includegraphics[width=9cm,height=5.5cm,angle=0]{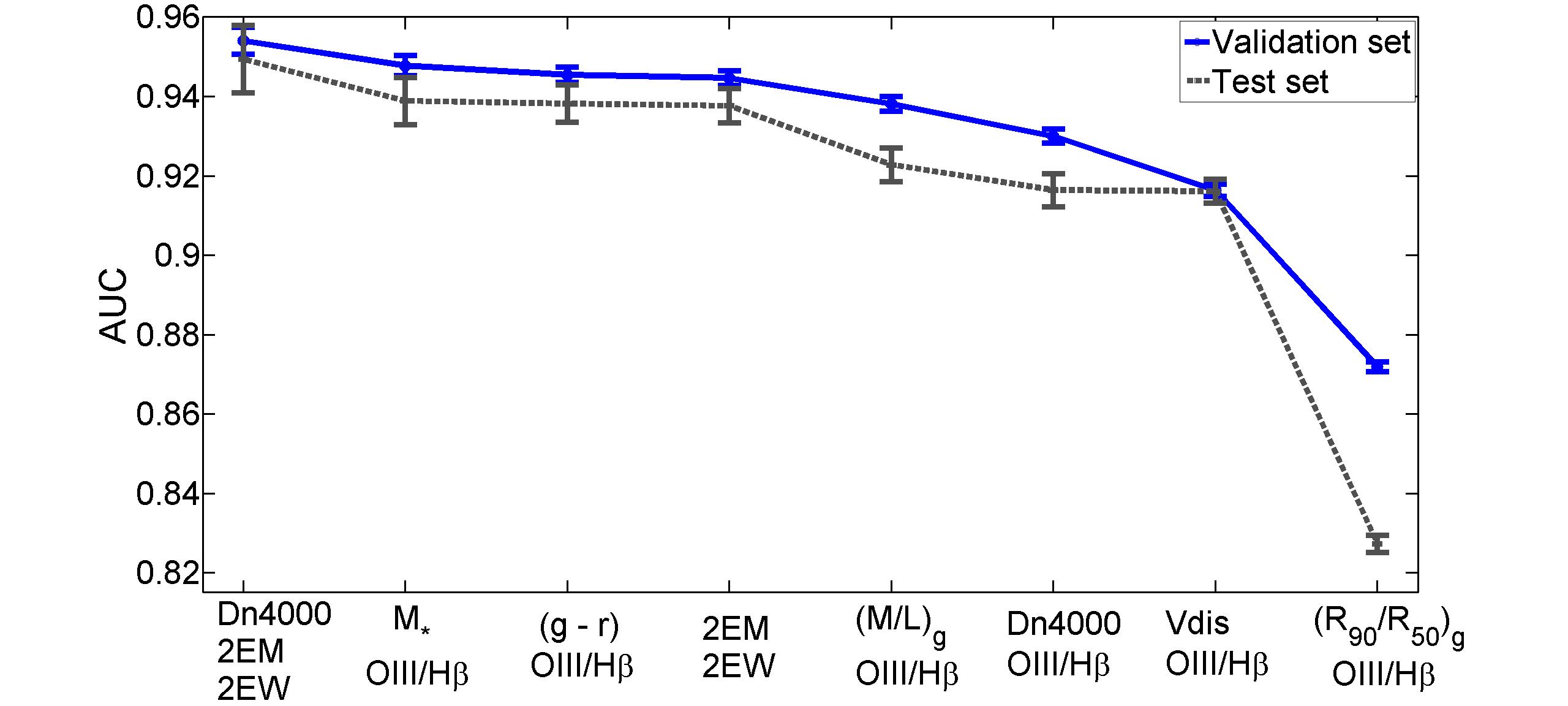}
\includegraphics[width=9cm,height=6.5cm,angle=0]{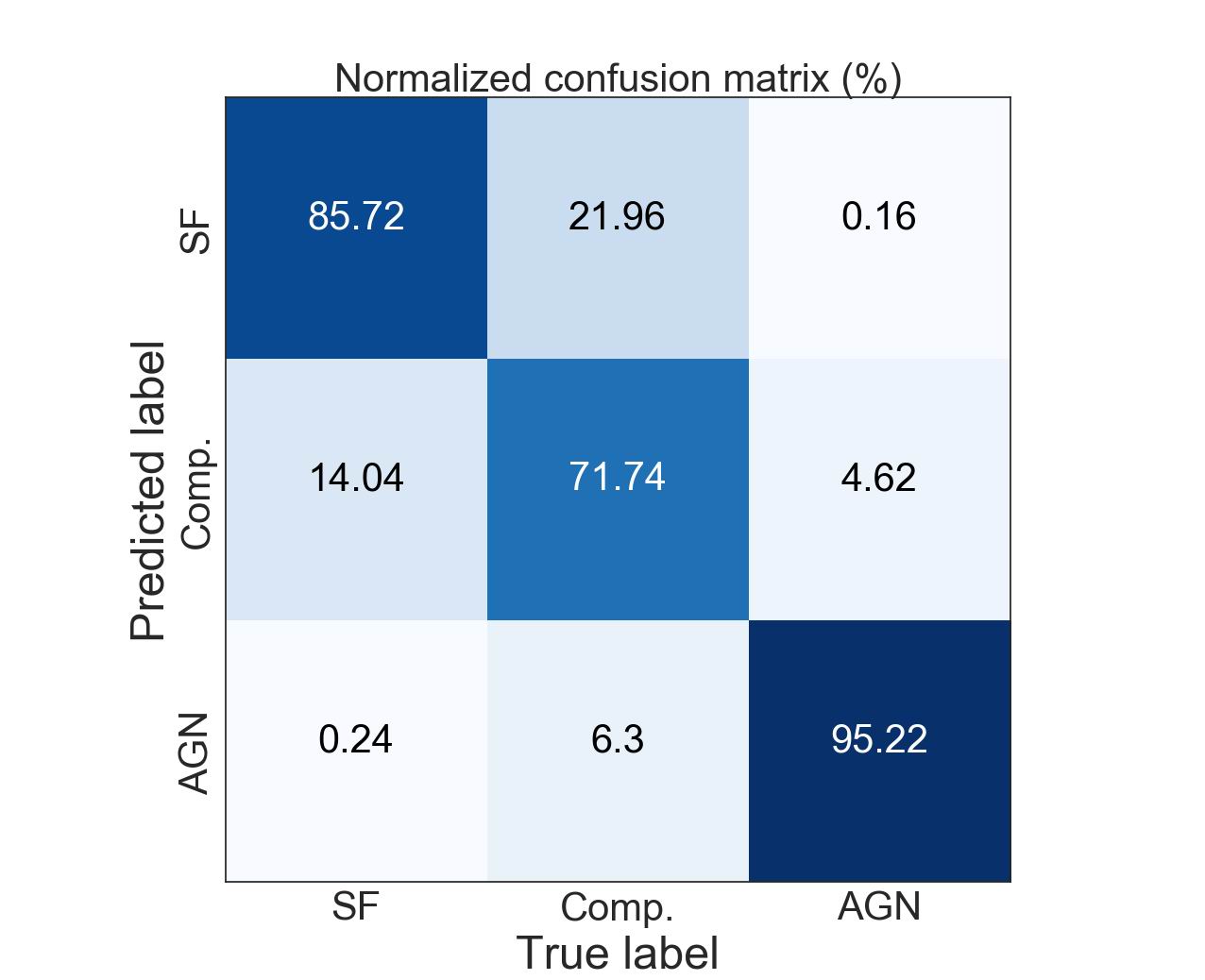}
\caption{The top plot shows the AUC for eight different combinations of the parameters used in this work.  The blue land grey lines represent the validation and test sets, respectively. The AUC values were calculated from 25 iterations of the training and classification process, after which the average and standard deviation are calculated from the top 20 AUC scores.  Here, Dn4000 along with the EW and flux of H$\beta$ and [OIII] contain the most informative data for classification. The stellar mass and g-r colour show the same ranks. The bottom plot shows the confusion matrix for the highest ranked parameter combination.}
\label{fig-auc-rank}
\end{figure}

The Dn4000 is an important parameter in the spectra of galaxies that can be added to the input for training the network. This parameter can be used to determine star formation rate in galaxies \citep{Brinchmann_2004} and it  also shows higher values for older galaxies.  We add this parameter to our training set and  in Figure \ref{fig-auc-rank},   summarize the average AUC produced by training the network using different combinations of physical and spectral parameters. The blue line represents the validation set with 30\% of the 15000 galaxies from the main training set. The gray dashed line represents the test set, which shows the same behavior as the validation set.  The AUC values were calculated from 25 iterations of the training and classification process, after which the average and standard deviation are calculated from the top 20 AUC scores \citep{Tem_2016}. Specifically, the order of the highest ranking parameter combinations remains the same between the two data sets. As can be seen, the combination of Dn4000 and the flux + EW of the H$\beta$ and [OIII] emission lines produces similarly high AUC values ($\sim$ 0.95) as stellar mass + [OIII]/H$\beta$ and g-r colour + [OIII]/H$\beta$. As shown in Figure \ref{fig-cm-xoh}, there are  considerable confusions, which are  mostly due to the SF galaxies. The star-forming galaxies in the bottom plot of Figure \ref{fig-cm-xoh} and in the red area under the S06 line, on average, have lower stellar mass with respect to the AGN galaxies above the K01 line.  Thus, the stellar mass can help remove the confusion between SF galaxies and AGN galaxies. This behaviour is also seen with the g-r parameter, for which AGN galaxies show a redder colour than the SF galaxies. Subtle differences in the correlations between the two EWs with mass and g-r do exist, however, which pattern recognition algorithms are capable of detecting in the underlying distributions.  In the ranking plot, we see that R$_{90}$/R$_{50}$ has the least relevant information out of all the parameters added to \ohb ~for this particular classification.

\begin{figure}
\centering
\includegraphics[width=9cm,height=5.5cm,angle=0]{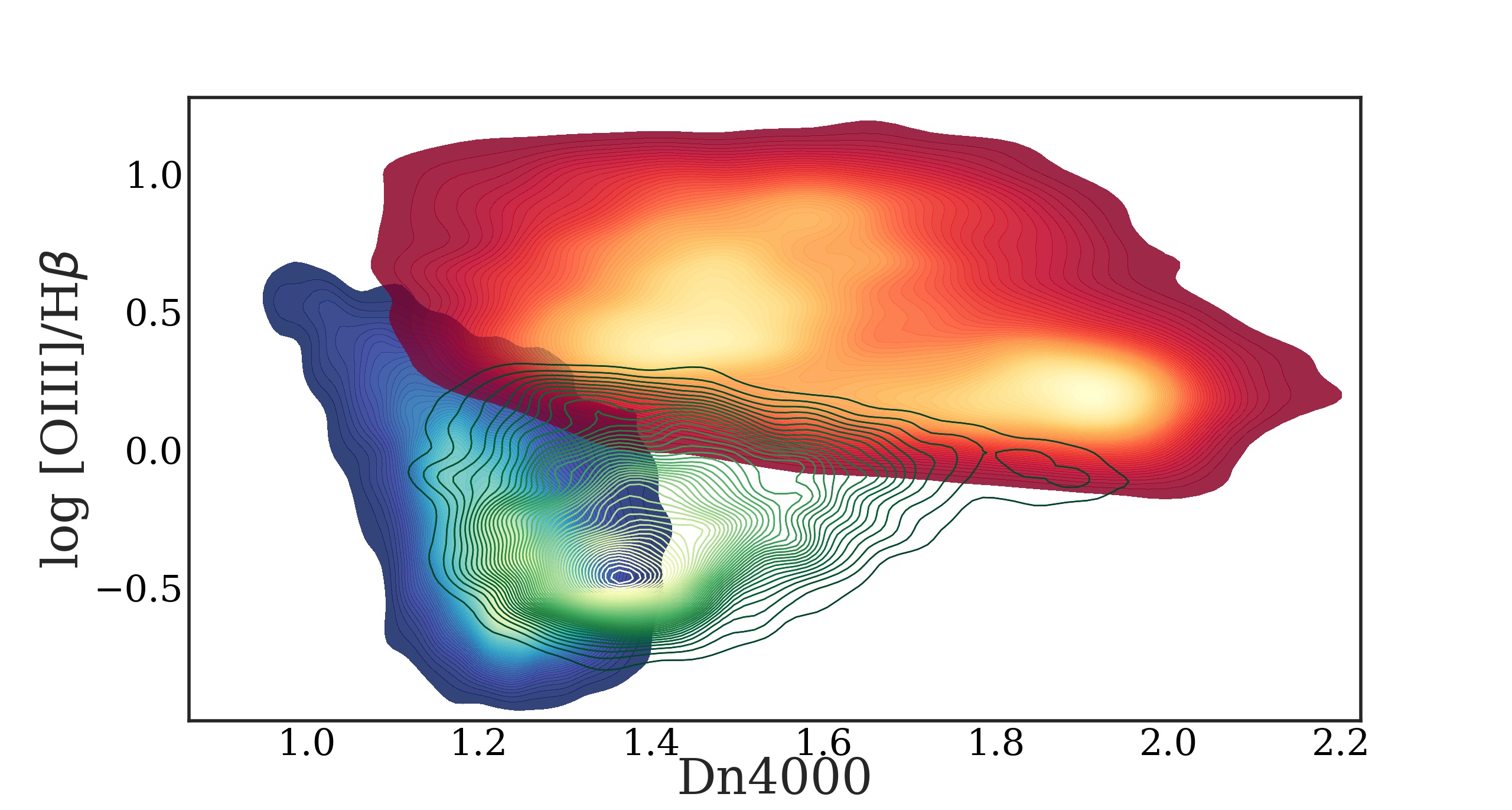}
\caption{The distribution of \ohb ~versus Dn4000 for the predicted classes. A combination of information in this plot and the two bottom  plots of Figure \ref{fig-cm-xoh4} are used to obtain a probability for a galaxy belonging to one of the three classes.}
\label{fig-bptp7-p7xoh4}
\end{figure}

In Figure \ref{fig-bptp7-p7xoh4} we show the distribution of \ohb\ versus Dn4000. A combination of different distributions, such as  the two bottom  plots of Figure \ref{fig-cm-xoh4}, are used to obtain a probability for a galaxy belonging to one of the three classes.  After obtaining the probabilities, a classification can be done based on a selected boundary decision \citep[see][]{Tem_2016}.  Generally, a decision boundary of 0.5 is a natural selection for a binary classification involving only two classes (i.e., if the probability that a galaxy belongs to a given class is above 50 \%, it is predicted to belong to that class).  For a problem involving three or more classes, however, the class with the highest predicted probability is usually identified as the predicted class.  All the confusion matrices presented for the three-class classification presented in this paper are based on the latter, `winning class' approach for determining the predicted class of each galaxy. 

\subsection{The error function }
\begin{figure*}
\centering
\includegraphics[width=8.cm,height=4.5cm,angle=0]{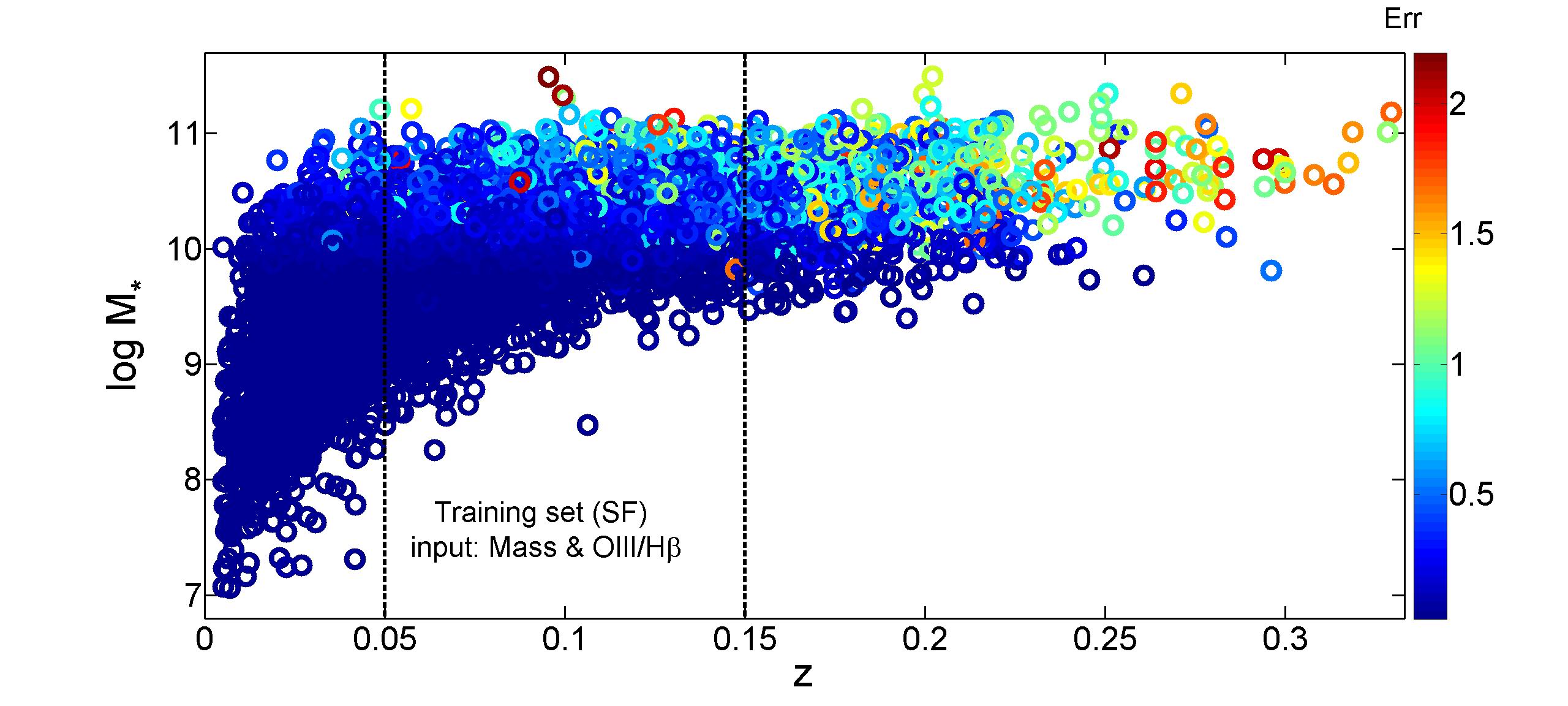}
\includegraphics[width=8cm,height=4.5cm,angle=0]{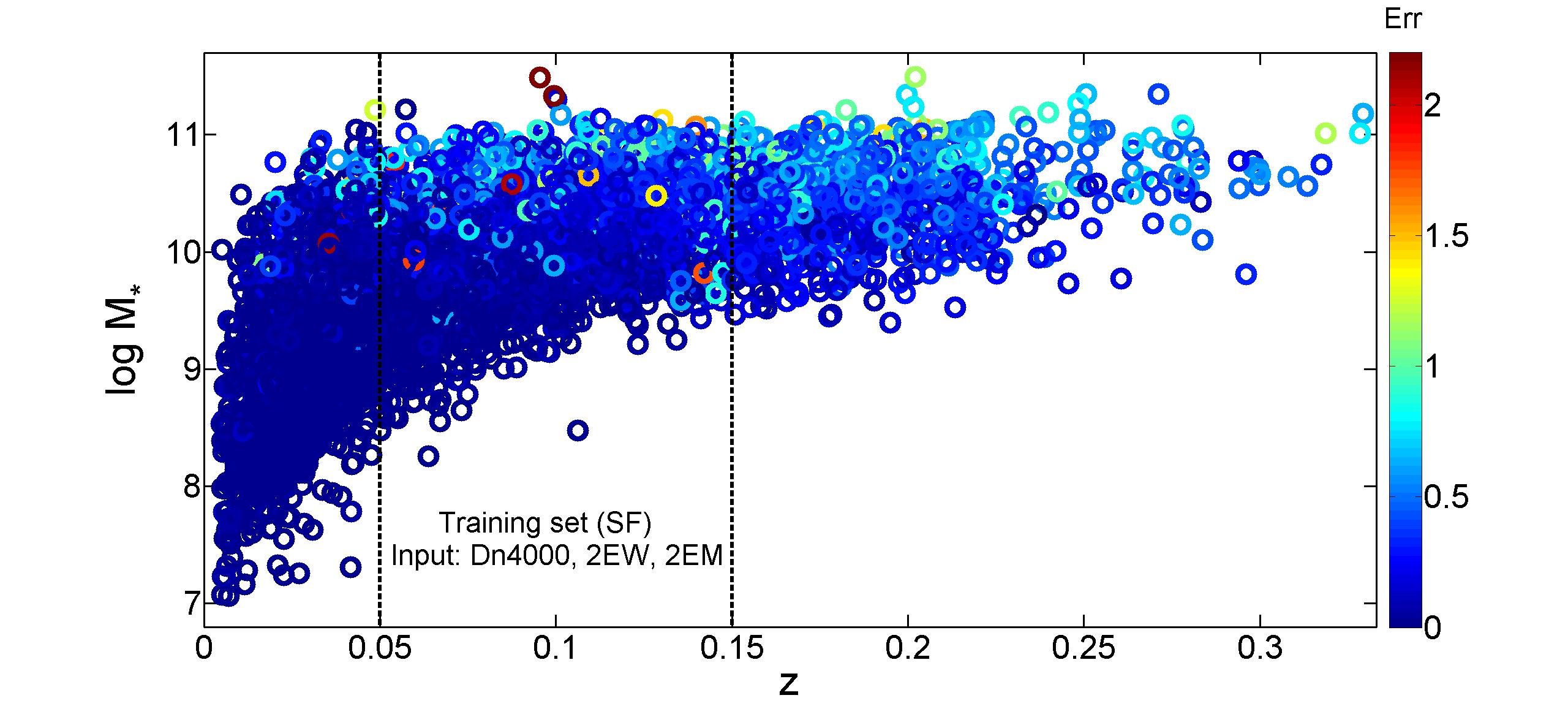}
\includegraphics[width=8cm,height=4.5cm,angle=0]{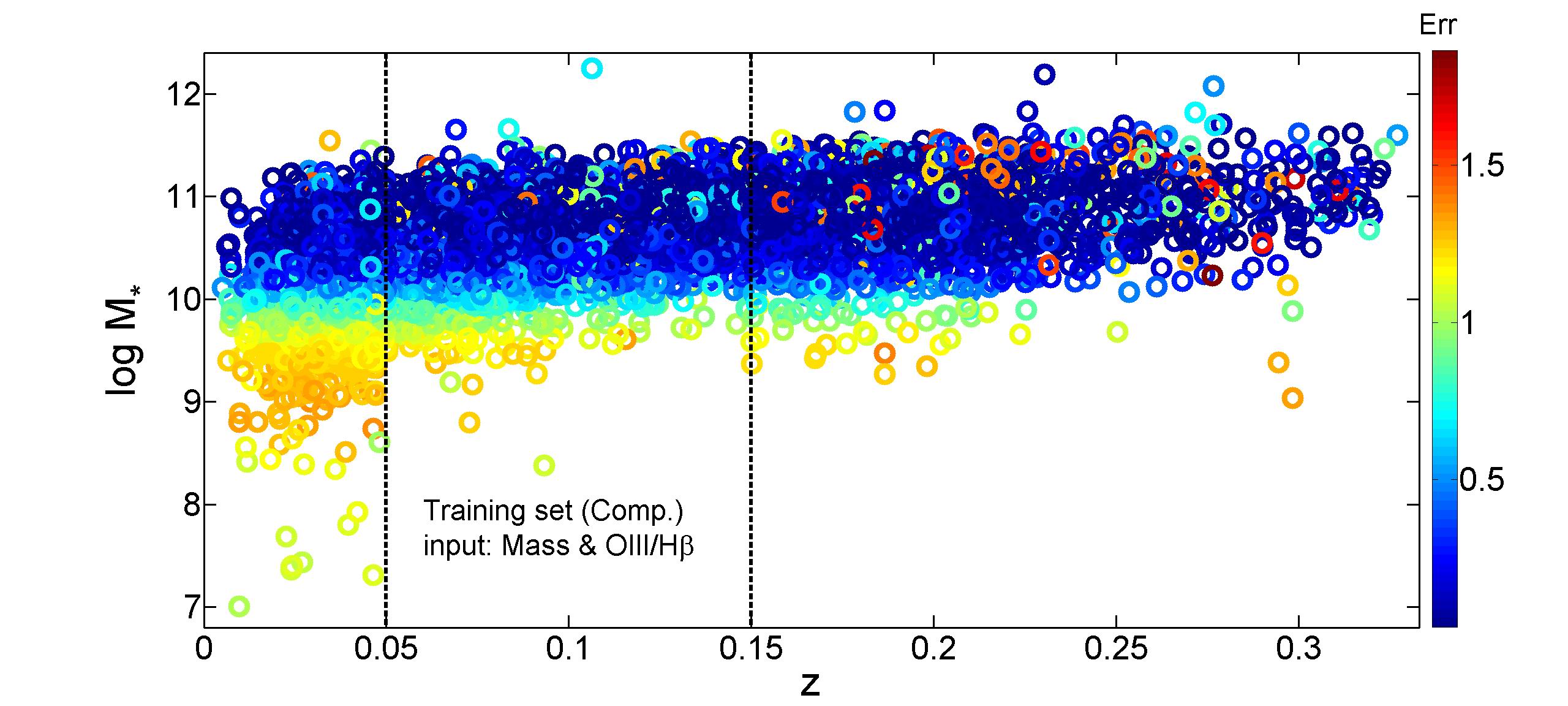}
\includegraphics[width=8cm,height=4.5cm,angle=0]{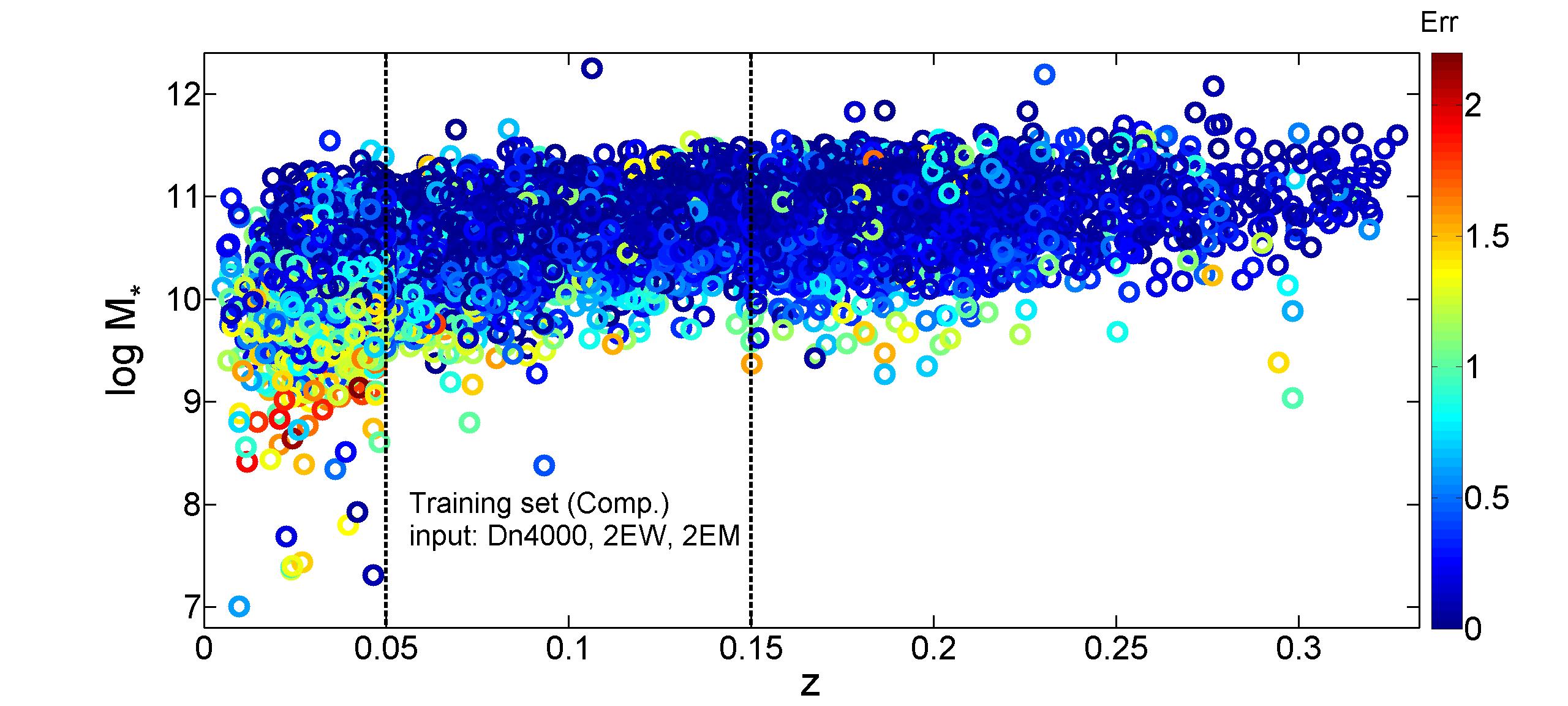}
\includegraphics[width=8cm,height=4.5cm,angle=0]{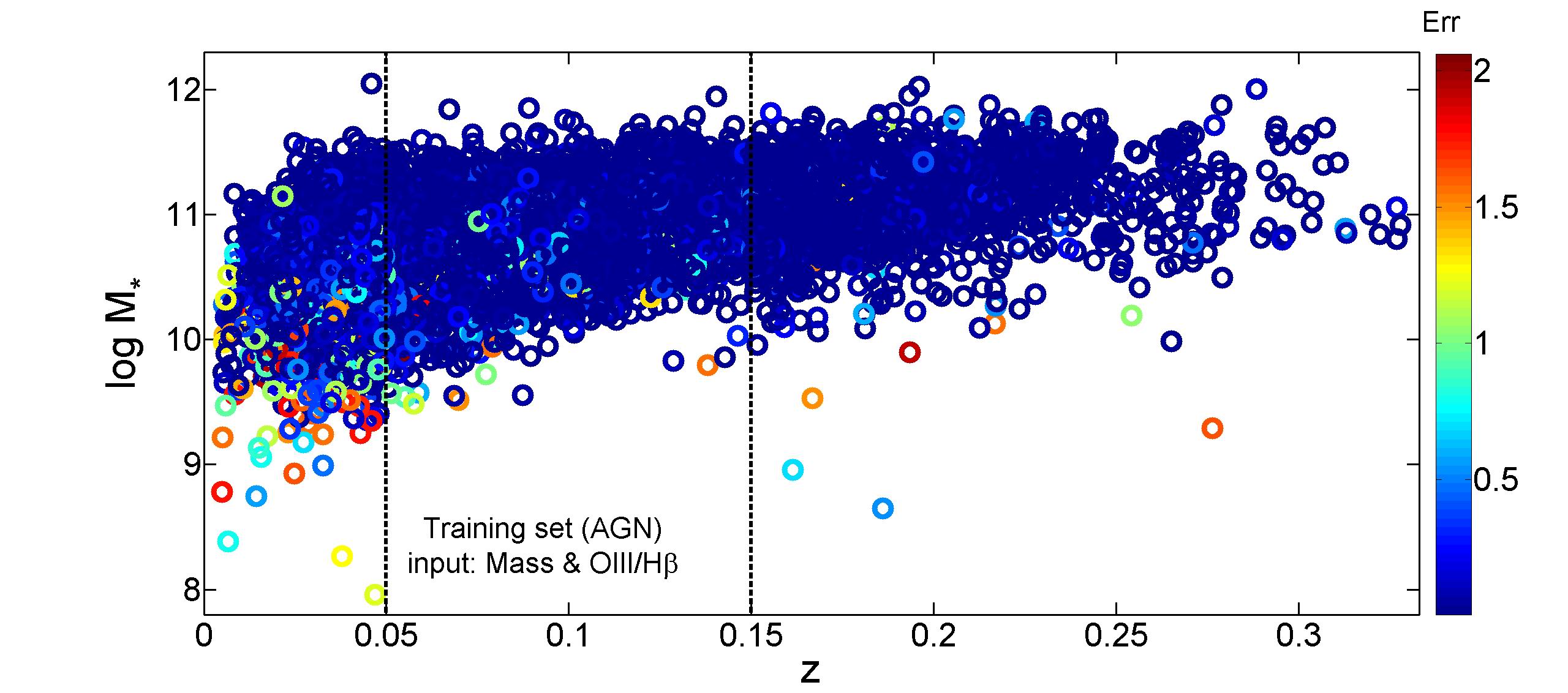}
\includegraphics[width=8cm,height=4.5cm,angle=0]{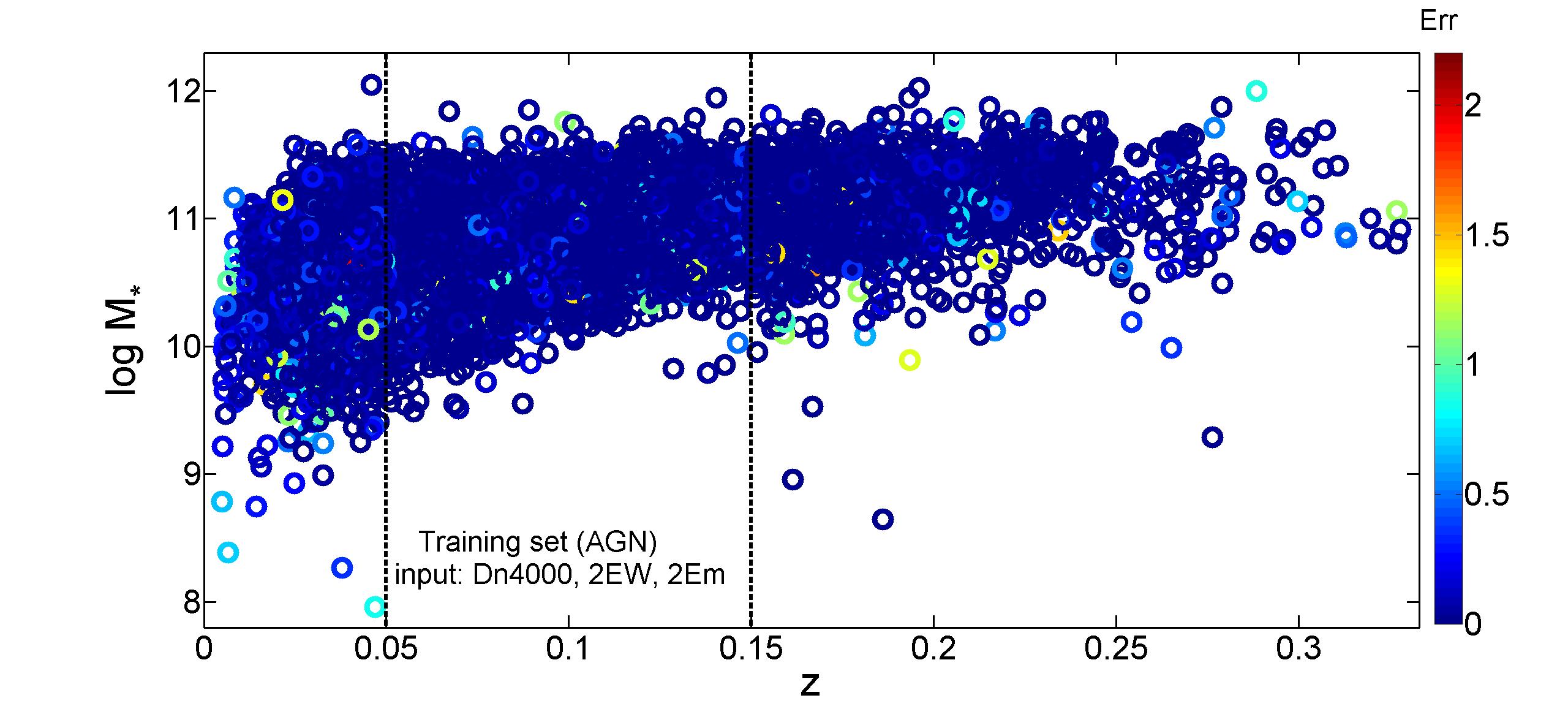}
\caption{ The stellar mass as function of redshift is shown, colour coded by the (cross-entropy) error. On the left panel, from top to bottom, we present the three classes of SF, composite, and AGN galaxies when the inputs are stellar mass and \ohb.  The galaxies between the two vertical lines are the main training set and  the rest of the galaxies are used as the test set. The right panel is the same as the left when we use Dn4000, the two emission lines and the two EWs as input (i.e., the best input predictors).}
\label{fig-zm-mass}
\end{figure*}

Since confusion matrices do not show a continuous behaviour of the predicted probabilities,  we  use a simple continuous error function as  $  \rm {Err = - log ~P}$, in which P is the predicted probability by ANN. This  error function  can be  useful for a (multi) binary classifications in which a true label of a each class is denoted by 1.  The error shows the deviation from a perfect prediction and is different from the error presented in Figure \ref{fig-auc-rank} (which is a statistical error). As an example,  for a perfectly predicted value of a class (P=1) we will have zero deviation from the true label  for that class. In Figure \ref{fig-zm-mass}, we show the stellar mass as a function of redshift, colour coded by the error. The error value of $\sim0.32$, for example, is related to P=0.5. So, the colours show lighter blue to red for the misclassified cases. In the left column, from top to bottom, we present the three classes of SF, composite, and AGN galaxies when the inputs are stellar mass and \ohb.  The galaxies between the two vertical lines in each panel represent the training set, while the rest of the galaxies are used as the test set. The right column is the same as the left when we use Dn4000, the two emission lines and the two EWs as input (i.e., the best input predictors presented in Figure \ref{fig-auc-rank}).  As can be seen, SF galaxies in the top left panel that have higher redshifts show higher errors.  This effect is less prominent for SF galaxies in the top right panel.  Likewise, the composite galaxies have a distinguishable error in the predictions for stellar mass $<10^{10} $ M$_\odot$ on the middle left panel.  In other words, when we use stellar mass as input, most of the confusion in composite galaxies occurs for the lower mass galaxies in this class. This behaviour also shows that composite galaxies with $\rm{log M_*>10}$ and \ohb\ can be predicted with high confidence.   It is noteworthy, however, that these effects are diluted in the plots in the right column, suggesting that predictions with Dn4000 plus the EWs and fluxes of [OIII] and H$\beta$ do not have  a strong dependency on the stellar mass.

\begin{figure*}
\centering
\includegraphics[width=8.cm,height=4.5cm,angle=0]{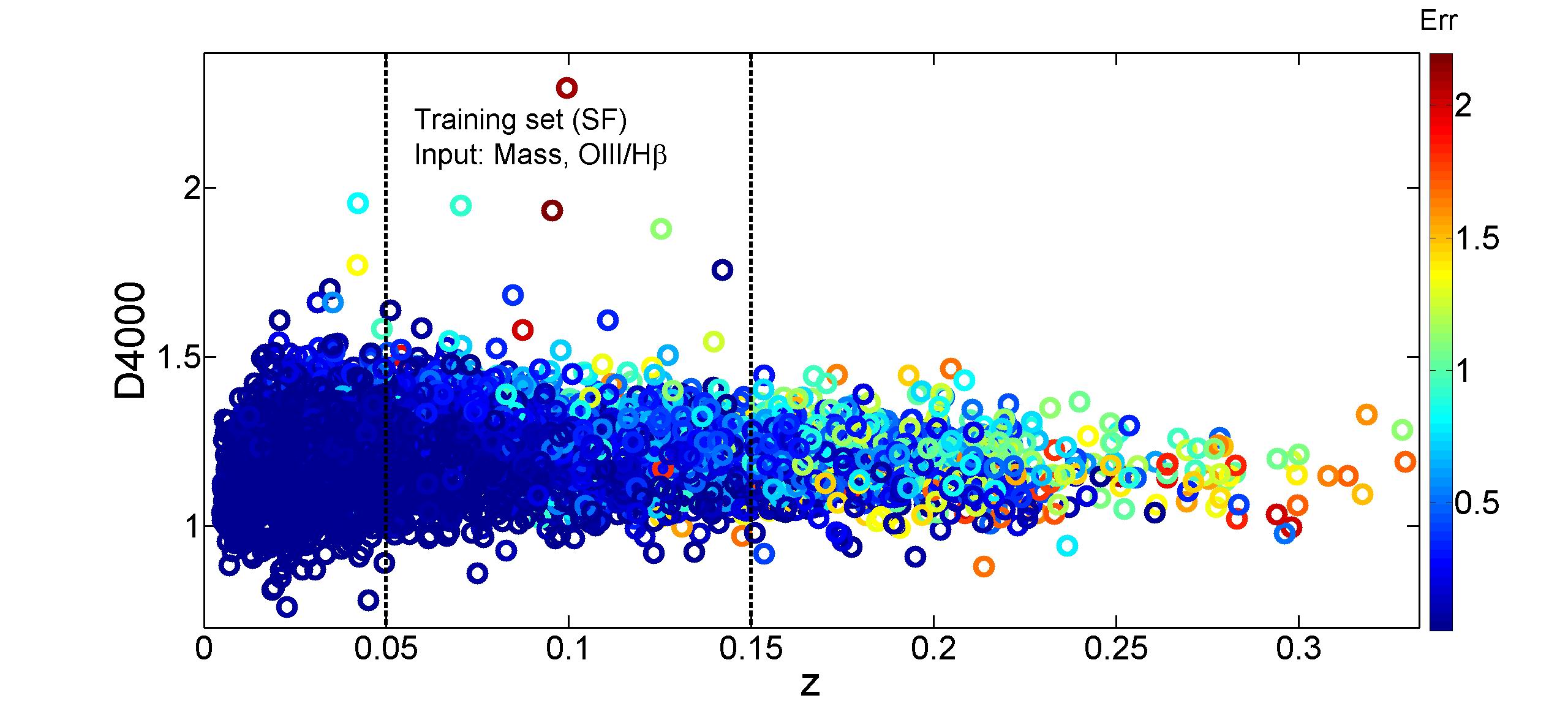}
\includegraphics[width=8cm,height=4.5cm,angle=0]{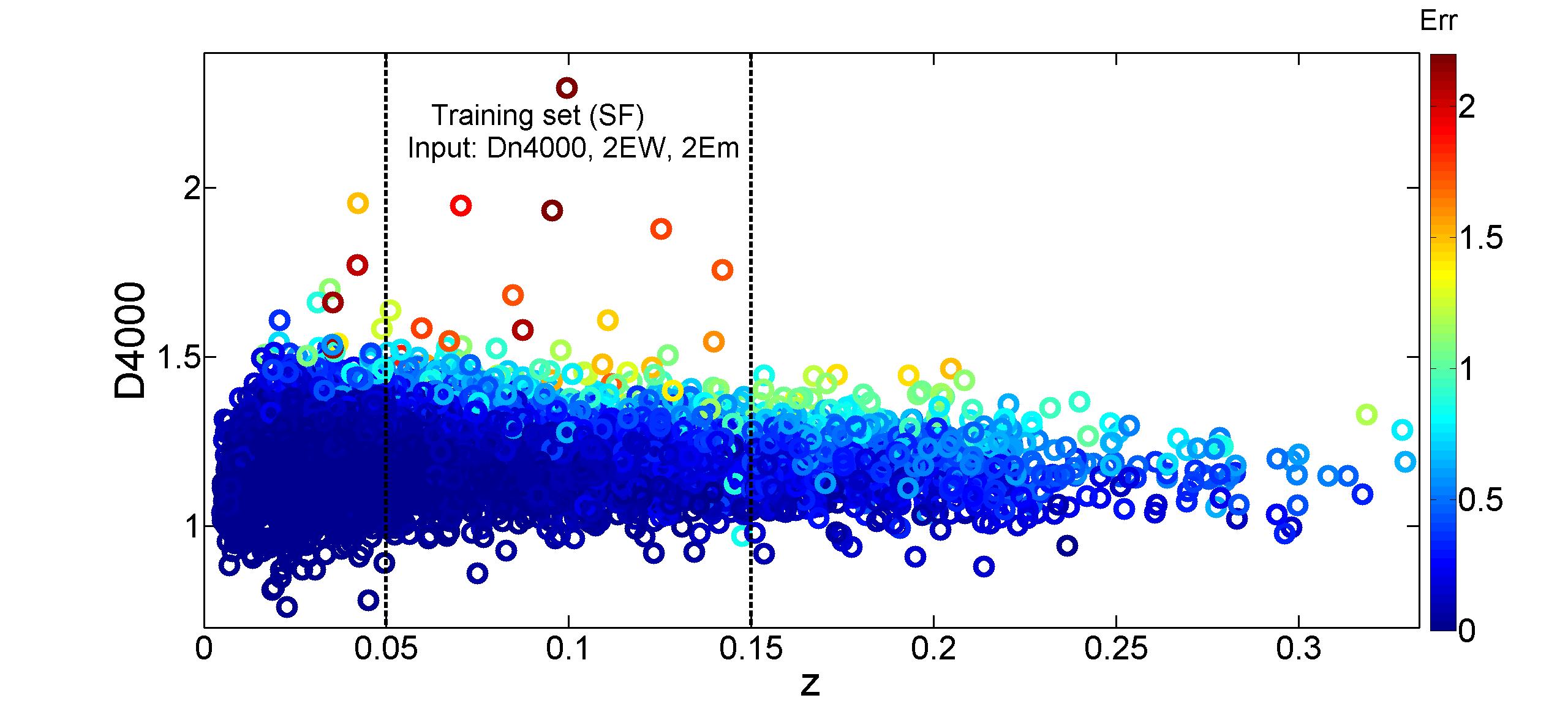}
\includegraphics[width=8cm,height=4.5cm,angle=0]{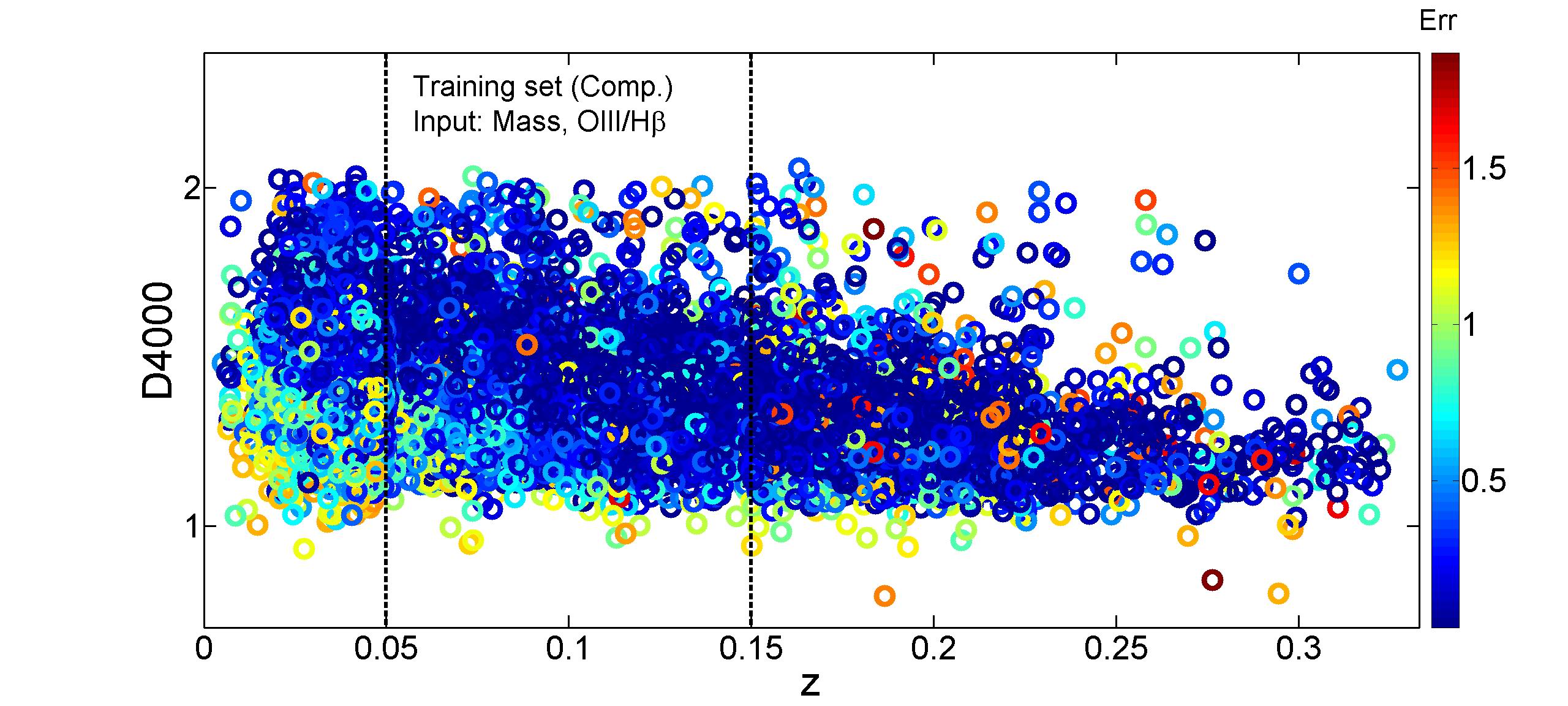}
\includegraphics[width=8cm,height=4.5cm,angle=0]{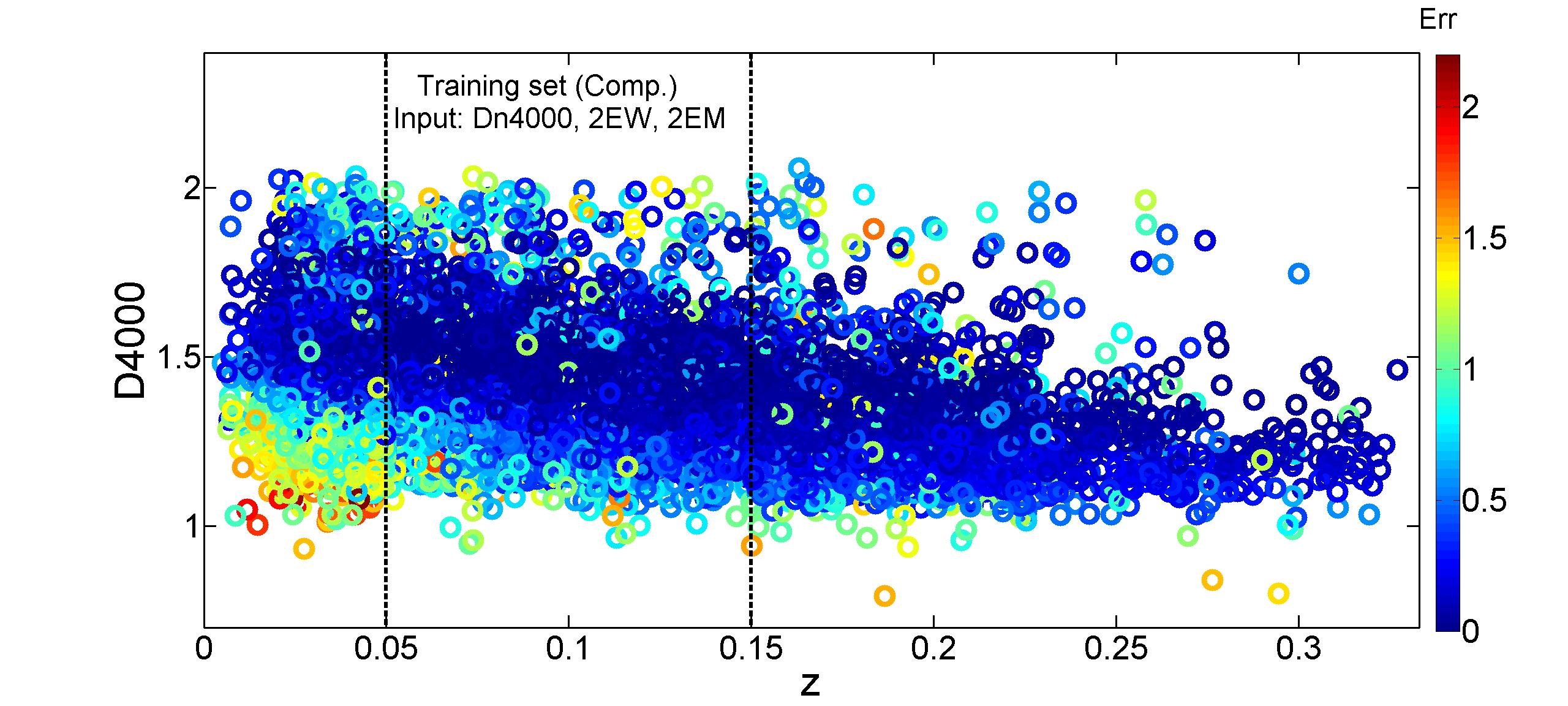}
\includegraphics[width=8cm,height=4.5cm,angle=0]{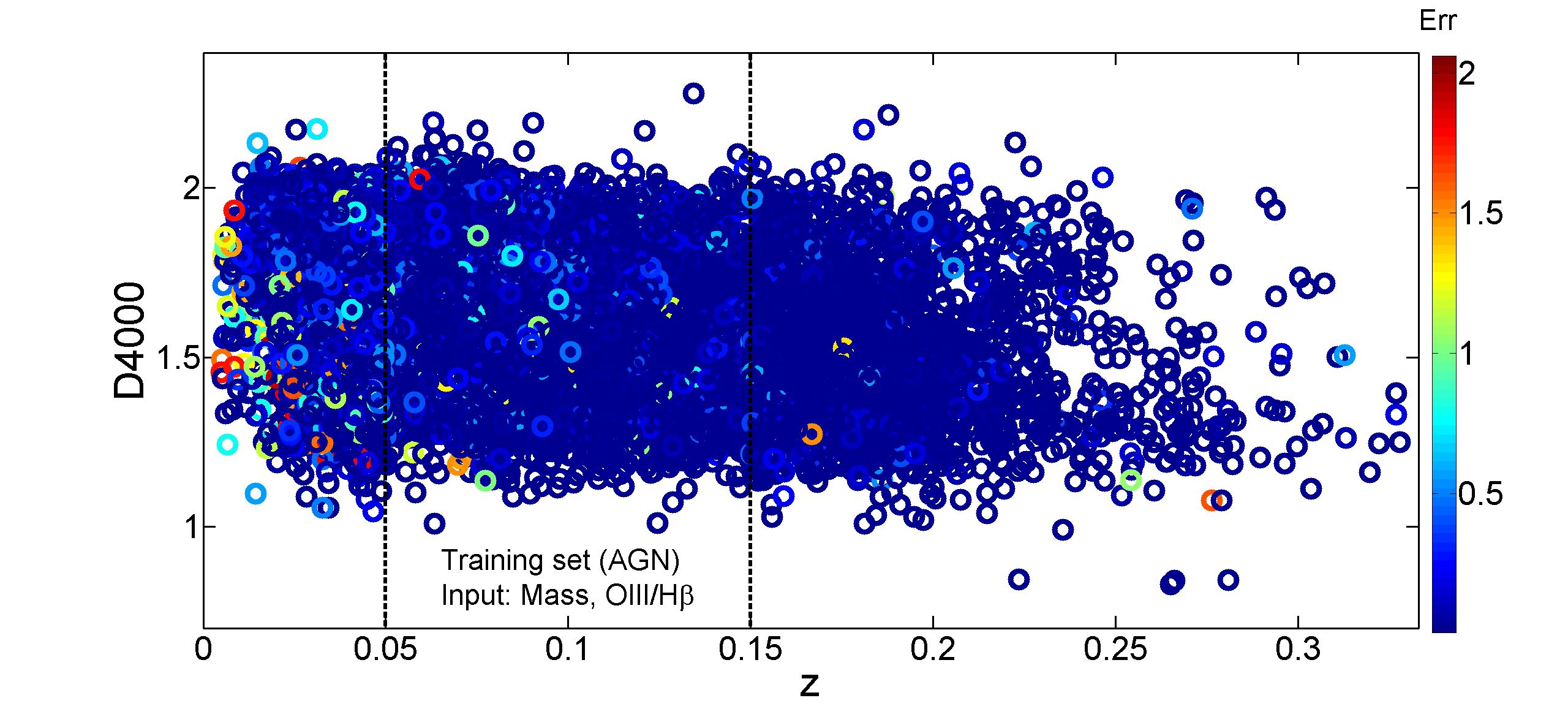}
\includegraphics[width=8cm,height=4.5cm,angle=0]{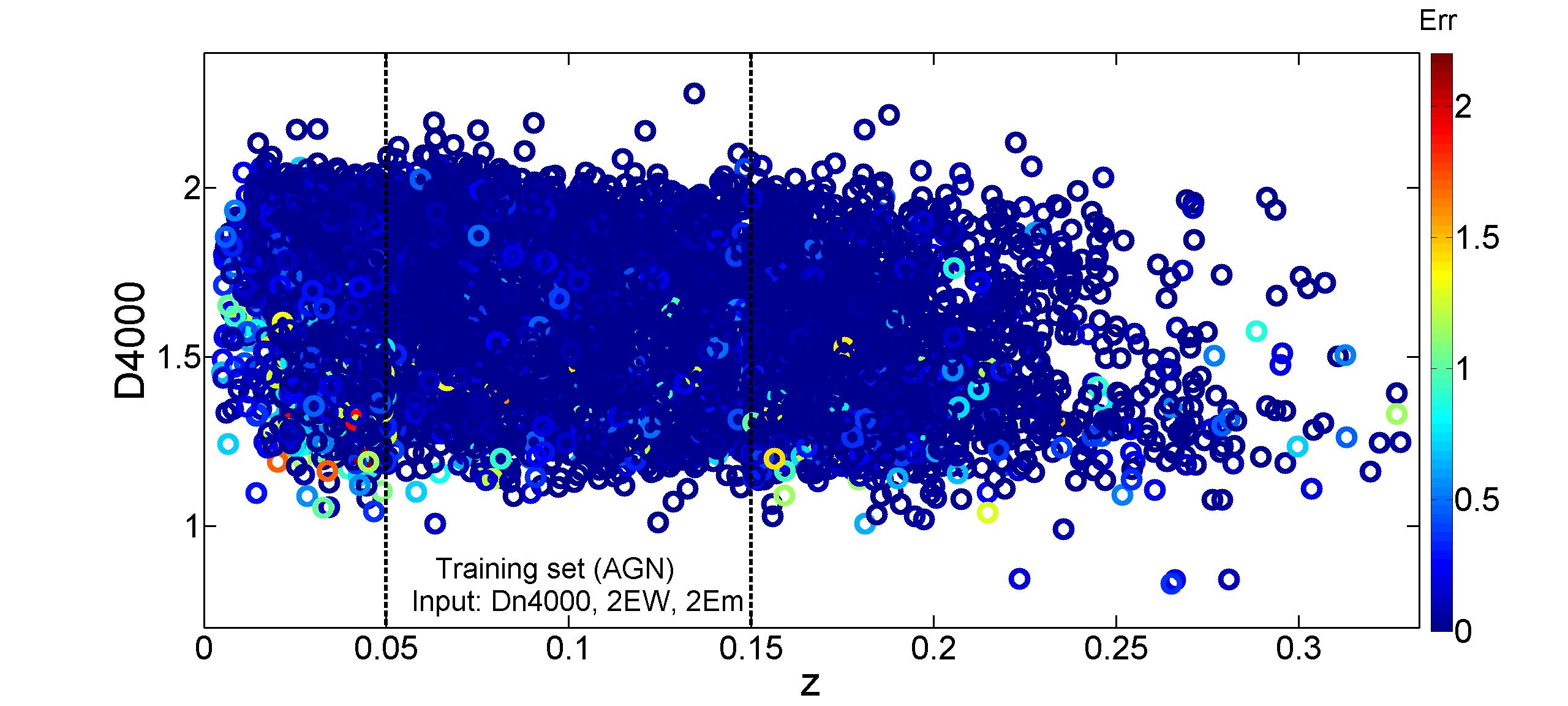}
\caption{This Figure is the same as Figure \ref{fig-zm-mass}, but for Dn4000 as function of redshift. The right panel is related to the best input predictors (shown in the top plot of Figure \ref{fig-auc-rank}) fed to the networks. The SF galaxies with higher values of Dn4000 show higher errors.}
\label{fig-zm-D4000}
\end{figure*}

Figure \ref{fig-zm-D4000} is similar to Figure \ref{fig-zm-mass}, but with Dn4000 instead of stellar mass plotted as a function of redshift.  The SF galaxies in the top left panel of this Figure show a dependency on redshift, which is not prominent in the top right panel when Dn4000 plus the EWs and fluxes of [OIII] and H$\beta$ are used as input.  The top right panel does, however, show that SF galaxies with unusually high values of Dn4000 are assigned high errors. Here, these galaxies do not show usual spectral behaviour, however and on average, some of their physical properties (which are not used in the related input) also show significantly higher values,  such as g-r ($\sim0.99\pm0.07$ mag) and Vdis ($\sim210\pm87$ km s$^{-1}$) with respect to the entire population (i.e., g-r $\sim0.45\pm0.14$ mag, Vdis $\sim77\pm36$ km s$^{-1}$). The average stellar mass also shows a higher value, however, it is not significant.

\subsection{A binary classification: Seyfert and LINERs}

\begin{figure}
\centering
\includegraphics[width=9cm,height=5.5cm,angle=0]{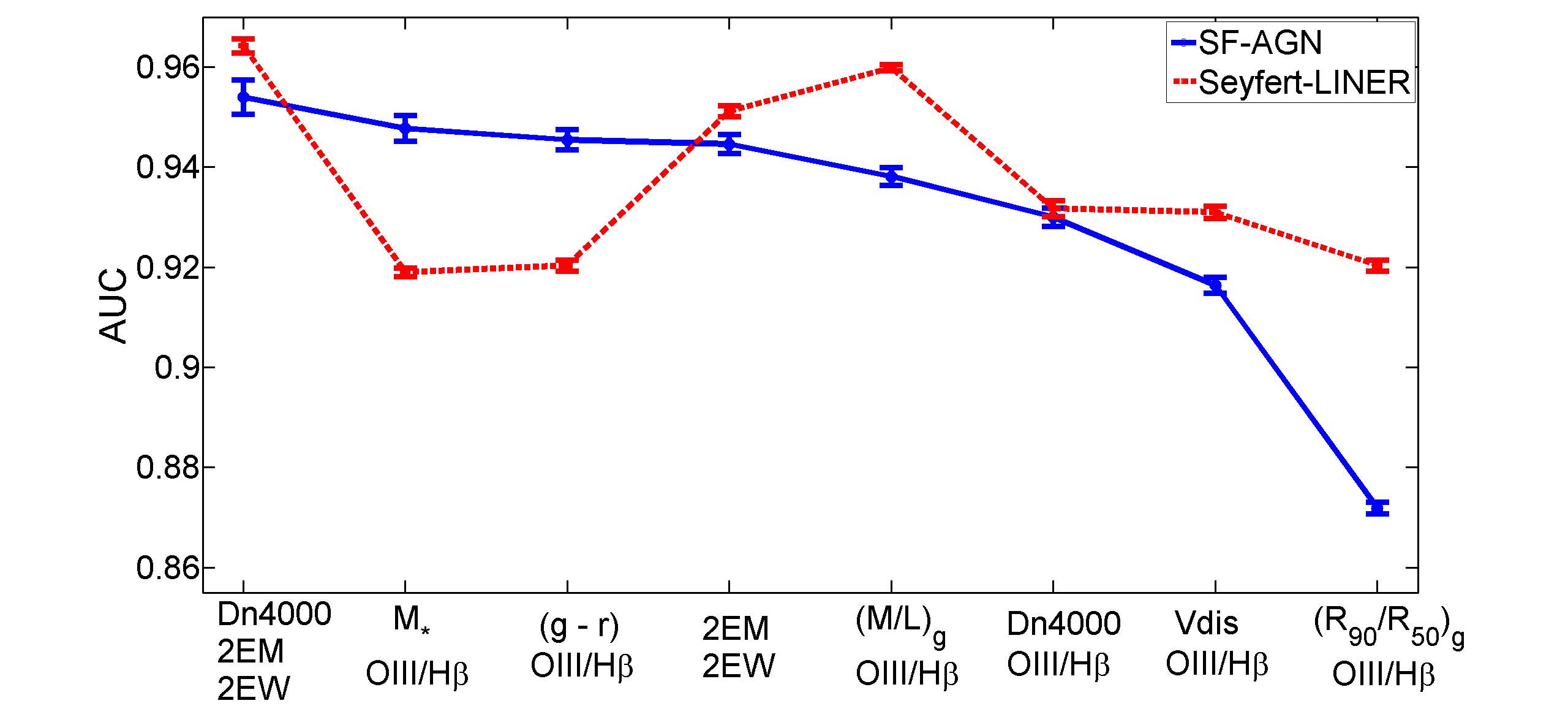}
\caption{The red dashed line shows the value of the AUCs produced for the Seyfert-LINER classification for the galaxies shown in the bottom plot of Figure \ref{fig-bpt-tr}.  The blue line is the three-class ranking presented in Figure \ref{fig-auc-rank}.  }
\label{fig-auc-rank2}
\end{figure}

The bottom plot of Figure \ref{fig-bpt-tr} is an example of a binary classification between Seyfert and LINER AGN galaxies.  We use the same method and input data presented in Figure \ref{fig-auc-rank} and show the resulting AUCs in Figure \ref{fig-auc-rank2}. In this plot, the blue  line represents the AUCs obtained from the three-class ranking presented in Figure \ref{fig-auc-rank}.  The red dashed line shows the result of the Seyfert-LINER classification. As can  be seen, Dn4000 along with the flux and EW of [OIII] and \hb\ provide the highest AUC. The stellar mass + [OIII]/H$\beta$, however, does not produce the second highest AUC, as was the case for the three-class classification.   In fact, stellar mass + [OIII]/H$\beta$ is the worst parameter combination in this Seyfert-LINER classification. On average, LINER galaxies generally have higher ages than Seyferts.  Since Dn4000 tends to be higher for an older population, a combination of \ohb\ and Dn4000 naturally provides better classifications. 

The second highest performing combination of parameters in the Seyfert-LINER binary classification includes the mass to light ratio.  M/L is dependent upon galaxy luminosity, with luminous galaxies showing high M/L while fainter galaxies show lower M/L over a broader range of values \citep{Kauffmann_2003}.  \cite{Kewley_2006} built upon this knowledge to show that Seyfert galaxies have slightly larger M/L values, on average, compared to LINERs.  The ANN trained in this paper has identified this slight difference in the M/L distributions between the two galaxy types and used it, in combination with [OIII]/H$\beta$, to provide efficient classifications.  Furthermore, \cite{Kauffmann_2003} also show that M/L is a sequence in Dn4000, with higher M/L correlating with higher values of Dn4000.  Thus, the high performance of M/L as a classifier for SDSS galaxies is likely related to the fact that Seyferts and LINERs represent younger and older populations, respectively, as discussed above for Dn4000.   

\section{Discussion}
\label{discussion}
\subsection{Comparison with previous classifiers}
Here, we compare our results to those of previous papers which have also attempted to use spectral and physical information to classify galaxies into SF versus AGN and/or Seyfert versus LINER.  For instance, Ju11 showed that stellar mass can be used to replace the [NII]/H$\alpha$ axis of the BPT diagram (their so-called `MEx' diagram) while still preserving the natural separation of SF, composite, and AGN galaxies.  Our network trained using stellar mass and [OIII]/H$\beta$ reproduces their result.  Specifically, placing the MEx diagram dividing line onto the bottom panel of Figure 5 would classify the galaxies in our sample in nearly the same way that our trained ANN has classified the galaxies.  

Our analysis goes beyond that of Ju11, however, because we also show that (g-r) colour can serve as an alternative replacement for [NII]/H$\alpha$ in the BPT diagram, providing equally efficient discriminatory power as stellar mass when classifying SDSS galaxies.  This result is consistent with the `CEx' method developed by \cite{Yan_2011}, which combines [OIII]/H$\beta$ with rest-frame (U-B) colour as classification parameters. Indeed, colour may actually be a superior choice over stellar mass because it does not require the additional assumptions and uncertainties involved in deriving stellar masses from broad-band spectral energy distribution fitting \citep{Mitchell_2013}.  

Similarly, we show that Dn4000  combined with the EWs and fluxes of [OIII] and H$\beta$ can provide even better discrimination between SF, composite, and AGN galaxies than stellar mass or (g-r) colour.  Not only is Dn4000 a better classifier than stellar mass for these galaxy classes, it also circumvents the additional uncertainties induced from stellar mass derivations since it is simply the ratio of the average flux density between two regions in the galaxy's spectrum \citep[e.g.,][]{Bruzual_1983, Balogh_1999}.  Thus, in surveys similar to the SDSS, using (g-r) colour or Dn4000 rather than stellar mass to separate SF, composite, and AGN galaxies provides equally efficient (or better) classifications with lower uncertainties.  

Ju11 show that the MEx diagram can also be used to separate AGN into Seyferts and LINERs.  They note, however, that there is considerable overlap between the two AGN sub-classes when using the MEx diagram.  Considering we found significantly worse classification performance for Seyferts and LINERs when using stellar mass as input, compared to the SF/composite/AGN classification performance also using stellar mass, our results echo the Ju11 conclusions.  The higher amount of flexibility that ANNs provide for exploring parameter space, however, allowed us to find that Seyferts and LINERs in the SDSS are best classified using Dn4000  in combination with their [OIII] and H$\beta$ emission line information.  Indeed, this example shows the advantage ANNs and other machine learning approaches have over traditional methods when there are many relevant parameters to consider.

K06 produced an alternative approach to separating Seyferts and LINERs which involved empirically-derived dividing lines overlaid onto the traditional BPT diagrams.  K06 also show that LINERs, on average, have older ages than Seyferts based on their Dn4000 and H$\delta$ absorption line measurements.  Our results confirm this conclusion since the top-ranked network for the Seyfert/LINER classification presented in this paper utilized Dn4000 as an input parameter.  Since the K06 dividing line for Seyferts and LINERs was used to determine the target classes for the network, however, the correspondence seen between our results and those of K06 is expected. 

Color-color diagnostics have also been found to be efficient Seyfert/LINER classifiers in previous studies.  For instance, \cite{Coziol_2015} used mid-infrared photometry from WISE to show that LINERs and Seyferts occupy distinct positions on a plane of W3-W4 versus W2-W3.  While our results show that (g-r) colour is surpassed by many other combinations of physical parameters for classifying Seyferts and LINERs, mid-infrared colours may provide improved discrimination between these galaxy types.  \cite{Coziol_2015} also suggest that the most probable explanation for such a dependence upon mid-infrared colour is related to an increasing rate of star formation from LINERs to Seyferts and up to SF galaxies.  Our results in the SF-AGN galaxy classification presented in this paper may support this claim, since we have shown that (g-r) colour can effectively discriminate SF galaxies from AGN galaxies.    

\subsection{Classifying Seyferts and LINERs on BPT diagram without H$\beta$}
As discussed in Section \ref{method} and shown in Figure 1, the target classes used to train the networks presented in this paper relied on H$\beta$ and [OIII] line flux measurements to place galaxies on the BPT diagram.  The input data we used to train the networks also included the H$\beta$ and [OIII] line flux measurements.  As such, one may argue that our strong discrimination between the galaxy classes when using the H$\beta$ line flux as an input to train the network is simply a function of our BPT classification scheme (i.e., the network was told each galaxy's class based on its H$\beta$ and [OIII] line flux).  To prove that these emission lines contain distinct patterns for each galaxy class, and that we are not biasing our results by using the H$\beta$ and [OIII] lines for target classifications, we conduct an alternative classification for Seyfert and LINERs in which H$\beta$ is not used to separate the two classes in the BPT diagram.  

\begin{figure}
\centering
\includegraphics[width=9cm,height=5.5cm,angle=0]{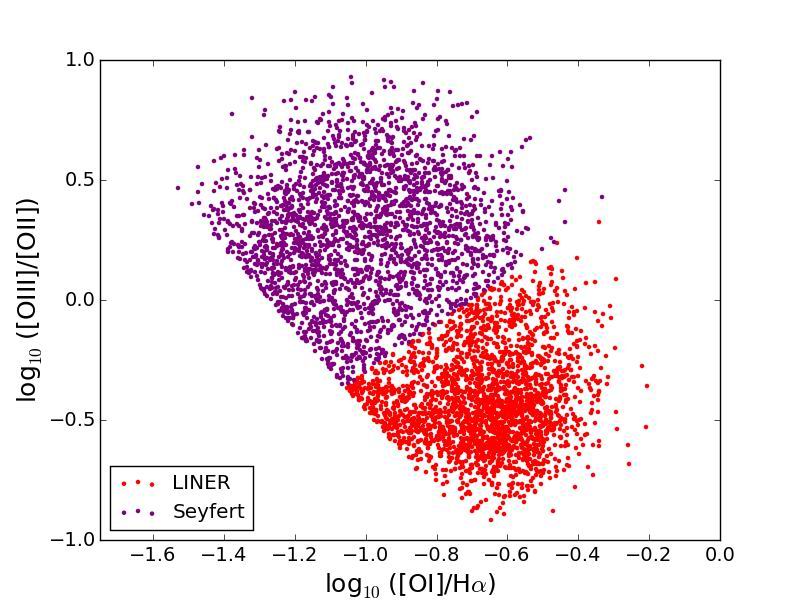}
\includegraphics[width=9cm,height=5.5cm,angle=0]{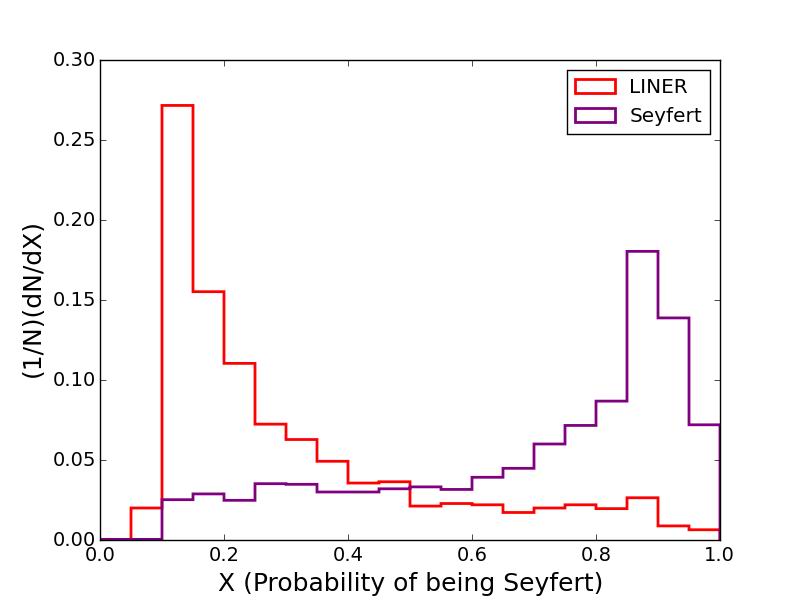}
\includegraphics[width=8cm,height=6.5cm,angle=0]{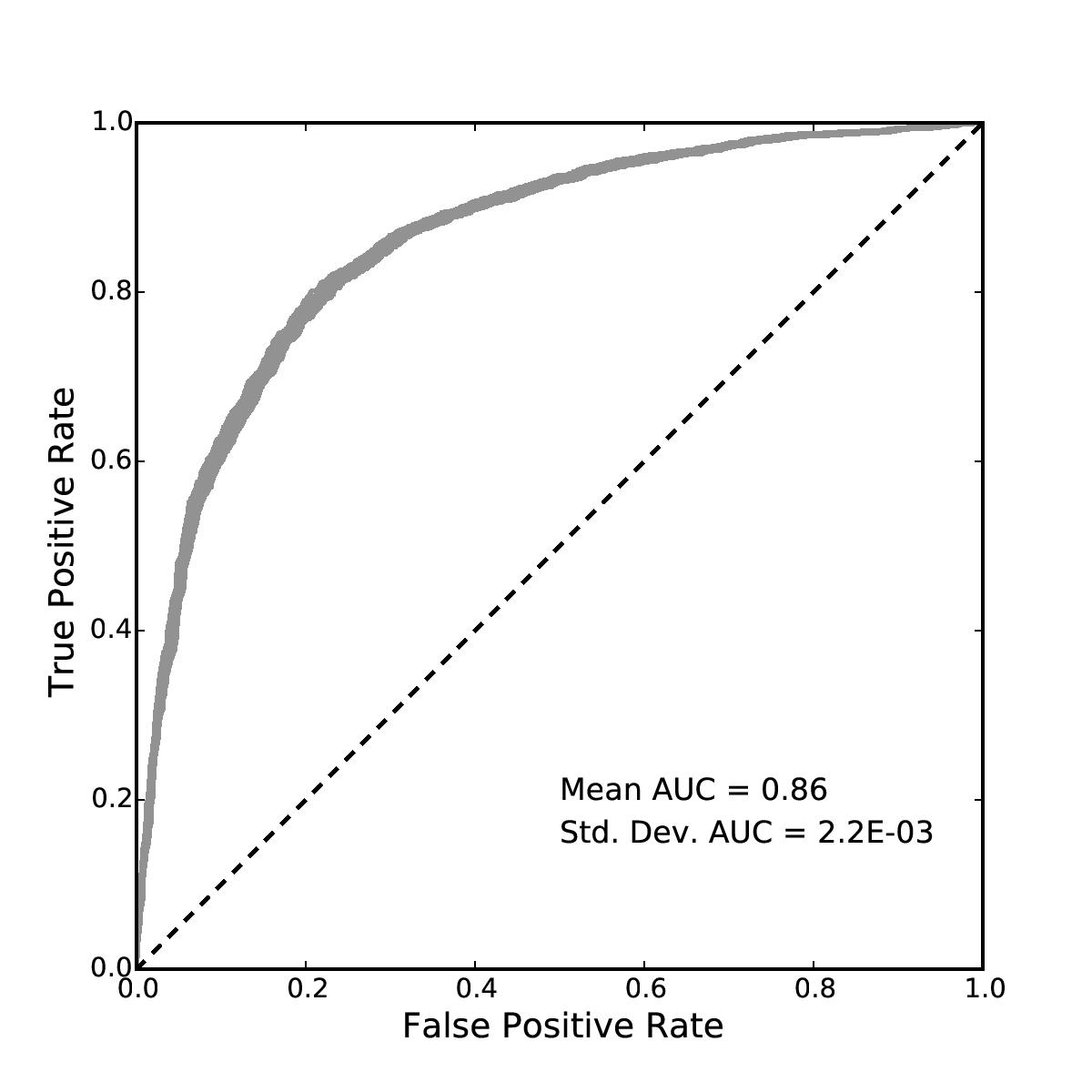}
\caption{The top panel shows the classification of 5000 AGN as LINERs and Seyferts based on their [OIII], [OII], [OI] and H$\alpha$ emission line fluxes and the empirical separation criteria developed by K06.  These classifications are used as targets to train a network to predict an AGN's class based solely on its H$\beta$ EW.  The results from the trained network are shown in the middle panel, which plots the probability distribution that a given galaxy is a Seyfert as predicted by the network for all the 5000 input galaxies.  The vertical axis has been normalized based on the total number of input galaxies.  The color of each histogram represents the `true' class of the galaxies based on their position in the top panel of this Figure.  The bottom panel shows the ROC plot for this binary classification.  The mean AUC and standard deviation were found by re-training the network using 50 different training sets drawn randomly from our full galaxy sample.}
\label{fig-roc-Sey2}
\end{figure}

K06 show that Seyferts and LINERs can be separated in a diagram of [OIII]/[OII] versus [OI]/H$\alpha$ line flux.  Here, we use the empirically derived Seyfert/LINER dividing line determined by K06 to determine a new Seyfert/LINER training set that is determined without the H$\beta$ line.  The top panel of Figure \ref{fig-roc-Sey2} shows the 5000 AGN galaxies (2500 from each class) with SNR $>$ 3 in the [OIII], [OII] (3726$\AA$), [OI] (6300$\AA$), and H$\alpha$ emission lines that were drawn randomly from our full sample and classified as Seyferts and LINERs based on the K06 criteria.  These classes were used as targets for a network that was trained using only H$\beta$ EW as input.  The middle panel of Figure \ref{fig-roc-Sey2} shows the probability distributions predicted by the trained network for the input galaxies based on their H$\beta$ EW.  The red histogram represents the distribution for the `true' LINERs, while the purple histogram represents the distribution for the `true' Seyferts.  The clear separation between the two classes shows that the network has correctly predicted a high probability of being a Seyfert for the majority of the `true' Seyferts.  It also correctly identifies the majority of the LINER population, assigning them low probabilities for belonging to the Seyfert class.  

The bottom panel of Figure \ref{fig-roc-Sey2} shows the ROC plot after training the network using 50 different training sets of 5000 AGN (requiring 2500 in each class for each iteration) using only H$\beta$ EW as input.  The 50 output ROC lines are plotted, along with the mean (0.86) and standard deviation (2.2$\times$10$^{-3}$) for the AUC.  Our results show that H$\beta$ provides relevant information that discriminates Seyfert and LINER galaxies, regardless of whether or not it is used to determine the `true' galaxy classes used as targets for the network.  Thus, using H$\beta$ flux to determine the `true' galaxy classes and as separate input data to train the network likely does not interfere with the conclusions drawn from the analysis presented in Section \ref{result}.  

\section{Conclusions}
\label{conclusion}
We have used artificial neural networks to show that star-forming, AGN, and composite galaxies can be discriminated based solely on a combination of physical and optical spectral measurements from the Sloan Digital Sky Survey that do not require near-infrared spectroscopy.  Our training set includes 15000 galaxies (5000 from each class determined by their position on the BPT diagram) selected randomly from a full sample of 160922 galaxies within the Max Planck Institute for Astrophysics (MPA)/Johns Hopkins University (JHU) galaxy catalogs.  After training multiple networks using various combinations of input parameters, we find that the fluxes and EWs of [OIII] and H$\beta$, combined with the Dn4000, provide the best segregation between the three galaxy classes.  Combining stellar mass or g-r colour with the [OIII] and H$\beta$ fluxes also produces a high rate of correct classifications.  Our method not only provides a way to classify galaxies in the absence of near-infrared spectra, but also requires only two optical emission lines to be measured at high SNR.

Additionally, we show that Seyfert and LINER galaxies are best classified by their Dn4000 plus their [OIII] and H$\beta$ fluxes and EWs.  Unlike the results from the star-forming/AGN/composite classification, however, stellar mass and g-r colour served as poor classifiers for the Seyfert/LINER classification.  This behaviour is likely due to the larger similarities in stellar mass and colour that exist between Seyfert and LINERs when compared to the more distinct differences between AGN and star-forming galaxies using those parameters.

\section*{Acknowledgments}
We sincerely thank Sara L. Ellison and James Di Francesco for their helpful discussions on this work and valuable comments that have improved this paper.  JK acknowledges the financial support of a Discovery Grant from NSERC of Canada.

\bibliographystyle{mnras}
\bibliography{HT-Jun2017}




\end{document}